\documentclass[fleqn,usenatbib]{mnras}

\usepackage{newtxtext,newtxmath}
\usepackage[T1]{fontenc}

\DeclareRobustCommand{\VAN}[3]{#2}
\let\VANthebibliography\thebibliography
\def\thebibliography{\DeclareRobustCommand{\VAN}[3]{##3}\VANthebibliography}

\usepackage{graphicx}	% Including figure files
\usepackage{amsmath}	% Advanced maths commands
\usepackage{threeparttable}

\newcommand{\oiii}{[\ion{O}{iii}]}	% [OIII]5007

\title[MRS spectroscopy of cid346]{JWST MIRI/MRS observations of hot molecular gas in an AGN host galaxy at Cosmic Noon}

\author[D. Kakkad et al.]{
D. Kakkad$^{1,2},$\thanks{E-mail: darshankakkad@gmail.com},
V. Mainieri$^{3}$,
Takumi S. Tanaka$^{4,5,6}$,
John D. Silverman$^{4,5,6,7}$,
D. Law$^{2}$,
Rogemar A. Riffel$^{8}$, \newauthor
C. Circosta$^{9,10}$,
E. Bertola$^{11}$,
M. Bianchin$^{12}$,
M. Bischetti$^{13,14}$,
G. Calistro Rivera$^{15}$,
S. Carniani$^{16}$,
C. Cicone$^{17}$,\newauthor
G. Cresci$^{18}$,
T. Costa$^{19}$,
C. M. Harrison$^{19}$, 
I. Lamperti$^{20,18}$,
B. Kalita$^{21,22,23}$,
Anton M. Koekemoer$^{2}$,\newauthor
A. Marconi$^{18,20}$,
M. Perna$^{24}$,
E. Piconcelli$^{25}$,
A. Puglisi$^{26}$,
Gabriele S Ilha$^{3,27}$,
G. Tozzi$^{28}$,
G. Vietri$^{29}$,\newauthor
C. Vignali$^{30,31}$,
S. Ward$^{32}$,
G. Zamorani$^{31}$,
L. Zappacosta$^{25}$,
\\
$^{1}$Centre for Astrophysics Research, University of Hertfordshire, Hatfield, AL10 9AB, UK\\
$^{2}$Space Telescope Science Institute, 3700 San Martin Drive, Baltimore, 21218,USA\\
$^{3}$European Southern Observatory, Karl-Schwarzschild-Strasse 2, Garching bei M\"{u}nchen, Germany\\
$^{4}$Department of Astronomy, Graduate School of Science, The University of Tokyo, 7-3-1 Hongo, Bunkyo-ku, Tokyo, 113-0033, Japan\\
$^{5}$Kavli Institute for the Physics and Mathematics of the Universe (WPI), The University of Tokyo Institutes for Advanced Study, Kashiwa, Chiba 277-8583, Japan\\
$^{6}$Center for Data-Driven Discovery, Kavli IPMU (WPI), UTIAS, The University of Tokyo, Kashiwa, Chiba 277-8583, Japan\\
$^{7}$Center for Astrophysical Sciences, Department of Physics \& Astronomy, Johns Hopkins University, Baltimore, MD 21218, USA\\
$^{8}$Departamento de F\'{i}sica/CCNE, Universidade Federal de Santa Maria, 97105-900 Santa Maria, RS, Brazil\\
$^{9}$ESA, European Space Astronomy Centre (ESAC), Camino Bajo del Castillo s/n, 28692 Villanueva de la Ca\~{n}ada, Madrid, Spain\\
$^{10}$Institut de Radioastronomie Millimétrique (IRAM), 300 rue de la Piscine, 38400 Saint-Martin-d’Hères, France\\
$^{11}$INAF-Osservatorio Astrofisico di Arcetri, Largo E. Fermi 5, I-50125 Firenze, Italy\\
$^{12}$Department of Physics and Astronomy, 4129 Frederick Reines Hall, University of California, Irvine, CA 92697, USA\\
$^{13}$Dipartimento di Fisica, Universitá di Trieste, Sezione di Astronomia, Via G.B. Tiepolo 11, I-34131 Trieste, Italy\\
$^{14}$INAF—Osservatorio Astronomico di Trieste, Via G. B. Tiepolo 11, I–34131 Trieste, Italy\\
$^{15}$German Aerospace Center (DLR), Institute of Communications and Navigation, 82234 Wessling, Germany\\
$^{16}$Scuola Normale Superiore, Piazza dei Cavalieri 7, I-56126 Pisa, Italy\\
$^{17}$Institute of Theoretical Astrophysics, University of Oslo, PO Box 1029, Blindern, 0315, Oslo, Norway\\
$^{18}$INAF – Osservatorio Astrofisico di Arcetri, Largo E. Fermi 5, I-50125, Florence, Italy\\
$^{19}$School of Mathematics, Statistics and Physics, Newcastle University, Newcastle upon Tyne, NE1 7RU, UK\\
$^{20}$Dipartimento di Fisica e Astronomia, Università di Firenze, Via G. Sansone 1, I-50019, Sesto F.no (Firenze), Italy\\
$^{21}$Kavli Institute for Astronomy and Astrophysics, Peking University, Beijing 100871, People's Republic of China\\
$^{22}$Kavli IPMU (WPI), UTIAS, The University of Tokyo, Kashiwa, Chiba 277-8583, Japan\\
$^{23}$Centre for Data-Driven Discovery, Kavli IPMU (WPI), UTIAS, The University of Tokyo, Kashiwa, Chiba 277-8583, Japan\\
$^{24}$Centro de Astrobiología (CAB), CSIC–INTA, Cra. de Ajalvir Km. 4, 28850 – Torrejón de Ardoz, Madrid, Spain\\
$^{25}$INAF – Osservatorio Astronomico di Roma, Via Frascati 33, I-00040, Monte Porzio Catone, Italy\\
$^{26}$School of Physics and Astronomy, University of Southampton, Highfield SO17 1BJ, UK\\
$^{27}$Universidade de S\~{a}o Paulo, Instituto de Astronomia, Geof\'{i}sica e Ci\^{e}ncias  Atmosf\'{e}ricas, Rua do Mat\~{a}o 1226, CEP 05508-090 S\~{a}o Paulo, SP, Brazil\\
$^{28}$Max-Planck-Institut für Extraterrestrische Physik (MPE), Giessenbachstraße 1, D-85748, Garching, Germany\\
$^{29}$INAF - Istituto di Astrofisica Spaziale e Fisica cosmica Milano, Via Alfonso Corti 12, 20133 Milano\\
$^{30}$Dipartimento di Fisica e Astronomia ‘Augusto Righi’, Universit\'{a} degli Studi di Bologna, via P. Gobetti, 93/2, 40129 Bologna, Italy\\
$^{31}$INAF- Osservatorio di Astrofisica e Scienza dello Spazio di Bologna, via Piero Gobetti, 93/3, I-40129 Bologna, Italy\\
$^{32}$Center for Computational Astrophysics, Flatiron Institute, 162 Fifth Avenue, NY 10010, USA\\
}

\date{Accepted XXX. Received YYY; in original form ZZZ}

\pubyear{2024}

\begin{document}

\label{firstpage}
\pagerange{\pageref{firstpage}--\pageref{lastpage}}
\maketitle

% Abstract of the paper
\begin{abstract}
Active Galactic Nuclei (AGN) are believed to play a central role in quenching star formation by removing or destroying molecular gas from host galaxies via radiation-pressure driven outflows and/or radio jets. Some studies of cold molecular gas in galaxies at Cosmic Noon ($z\sim2$) show that AGN have less cold gas ($<$100 K) compared to mass-matched star-forming galaxies. However, cold gas could also be shock-heated to warmer phases, detectable via H$_{2}$ transitions in the rest-frame near- and mid-infrared spectra. The Medium Resolution Spectrograph (MRS) of the Mid-infrared Instrument (MIRI) aboard JWST has opened a unique window to observe these emission lines in galaxies at Cosmic Noon. We present the first detection of hot molecular gas in cid\_346, an X-ray AGN at $z\sim2.2$, via the H$_{2}$ ro-vibrational transition at 2.12 $\mu$m. We measure a hot molecular gas mass of $\sim 8.0 \times 10^{5}$ M$_{\odot}$, which is $\sim 10^{5}-10^{6}$ times lower than the cold molecular gas mass. cid\_346 is located in an environment with extended gas structures and satellite galaxies. This is supported by detection of hot and cold molecular gas out to distances $>$10 kpc in MIRI/MRS and ALMA data, respectively and ancillary NIRCam imaging that reveals two satellite galaxies at distances of $\sim$0.4 arcsec (3.3 kpc) and $\sim$0.9 arcsec (7.4 kpc) from the AGN. Our results tentatively indicate that while the CO(3-2)-based cold gas phase dominates the molecular gas mass at Cosmic Noon, H$_{2}$ ro-vibrational transitions are effective in tracing hot molecular gas locally in regions that may lack CO emission.
\end{abstract}

% Select between one and six entries from the list of approved keywords.
% Don't make up new ones.

%%%%%%%%%%%%%%%%%%%%%%%%%%%%%%%%%%%%%%%%%%%%%%%%%%

%%%%%%%%%%%%%%%%% BODY OF PAPER %%%%%%%%%%%%%%%%%%

\section{Introduction} \label{sect1}

% Things to check in the pipeline
% Check jum_threshold or jump_detection parameter or rejection_threshold

% H2 lines reference:
% https://www.mpe.mpg.de/ir/ISO/linelists/H2.html

Feedback from growing supermassive black holes (SMBHs), also known as Active Galactic Nuclei (AGN), has become a fundamental component of cosmological simulations that model galaxy formation and evolution \citep[e.g.,][]{hirschmann14, dubois16, genel14, schaye15, pillepich18, schaye23}. AGN feedback plays a key role in limiting star formation efficiency, particularly in galaxies with high halo masses \citep[e.g.,][]{benson03, somerville08, harrison17, rennehan24}. As a result of incorporating these feedback mechanisms, simulations now closely match the observed stellar mass functions of galaxies, especially at low redshifts \citep[see also][]{silk12}.

Observations of multi-phase gas outflows have been crucial in identifying signatures of ongoing AGN-driven feedback. Numerous studies, ranging from fibre or slit spectroscopy to integral field spectroscopy (IFS) in rest-frame optical wavelengths, have successfully detected fast winds (velocity, $v > 1000$ km s$^{-1}$) in the ionised gas phase of AGN, star-forming, and starburst galaxies. These studies often use \oiii$\lambda$5007 (\oiii~hereafter) or H$\alpha$ as ionised gas tracers, covering galaxies from low-redshift to Cosmic Noon ($1<z<3$) to the Epoch of Reionisation \citep[e.g.,][]{arribas14, harrison14, forster-schreiber18, rakshit18, davies20, kakkad22, kakkad23a, travascio24, carniani24, loiacono24, roy24, tozzi24, vayner24}. Outflows have also been observed in the molecular gas phase, where they are believed to carry the bulk of the outflow mass in galaxies \citep[e.g.,][]{cicone14, carniani15, fiore17, fluetsch19, lutz20, veilleux20}. These outflows are expected to be especially important in delivering feedback into the interstellar medium (ISM) during Cosmic Noon, when both the volume-averaged star formation rate (SFR) and AGN accretion activity reach their peak \citep[e.g.,][]{shankar09, madau14}.

The relationship between AGN activity and star formation, particularly the impact of AGN-driven outflows on SFR, remains a topic of active debate in the literature. AGN activity is typically identified by excess emission in various wavelengths—beyond what can be attributed to star formation or internal processes within the host galaxy—such as hard X-rays, mid-infrared, or radio, as well as through nebular diagnostics in UV, optical, or infrared wavelengths \citep[e.g.,][]{baldwin81, hirschmann23}. The timescales of this AGN activity can differ by several orders of magnitude \citep[$> 10^{5}$ yrs:][]{schawinski15, padovani17}. SFR, on the other hand, is conventionally measured using short-term tracers ($\sim$10 Myr) like H$\alpha$ or [\ion{O}{ii}]$\lambda$3727 emission lines \citep[e.g.,][]{glazebrook99, murphy11, calzetti13, tacchella22}, or long-term tracers ($\sim$100 Myr) such as far-infrared (FIR) or sub-mm continuum \citep[see reviews by][]{kennicutt98, kennicutt12}. This difference in timescales between AGN activity and star formation measurements likely contributes to the wide range of findings on the AGN-SFR relationship in the literature. For example, \citet{rosario12} found that the connection between FIR luminosity (an SFR indicator) and AGN accretion rate depends on both AGN luminosity and redshift. At high redshifts, in particular, there appears to be no clear relationship between FIR-based SFR and AGN accretion activity. However, spatially resolved studies of H$\alpha$ emission in a few galaxies suggest that AGN outflows may either suppress or enhance star formation locally, on sub-kiloparsec to kiloparsec scales \citep[e.g.,][]{cano-diaz12, cresci15, carniani16, bessiere22, kakkad23b}, a result also supported by recent zoom-in simulations \citep[see][]{mercedes-feliz24}.

Molecular gas, however, offers a promising alternate avenue for assessing the impact of AGN feedback in galaxies. Since stars form from cold molecular gas, AGN-driven outflows, whether by radiation pressure or jets, are likely to impact the molecular gas first. These outflows could sweep up and eject the cold gas (`ejective' feedback), or increase the temperature or turbulence within the gas (`preventative' feedback). Recent simulations with EAGLE suggest that AGN host galaxies contain larger molecular gas reservoirs compared to star-forming galaxies. For example, \citet{ward22} investigated the molecular gas content of quasars in three cosmological simulations, namely EAGLE \citep{schaye15}, IllustrisTNG \citep{nelson19} and SIMBA \citep{dave19} and concluded that AGN host galaxies should have similar or enhanced molecular gas fractions compared to mass-matched inactive galaxies (or normal star forming galaxies), both at z=0 and z=2. However, observations present conflicting results. APEX-based CO observations of low-redshift, X-ray-selected AGN host galaxies \citep[e.g.,][]{koss21} support the idea that AGN often reside in gas-rich environments. In contrast, other studies report no significant difference in local molecular gas content between AGN hosts and mass-matched star-forming galaxies at low redshift \citep[e.g.,][]{rosario18, husemann17}, or even suggest that AGN host galaxies have lower molecular gas content in their central regions compared to the rest of the galaxy \citep[e.g.,][]{ellison21, garcia-burillo24}. At Cosmic Noon, follow-up campaigns using ALMA and NOEMA have indicated faster gas depletion in AGN hosts compared to star-forming galaxies \citep[e.g.,][]{kakkad17, bischetti21, circosta21, bertola24, frias-castillo24}.

The studies mentioned above predominantly use different excitations of the CO molecule as tracers for cold molecular gas \citep[see reviews by][]{carilli13, saintonge22}. This cold phase of molecular gas is at temperatures of T$\leq$100 K. But in turbulent environments with strong AGN radiation, gas can be excited to higher temperatures, a phenomenon extensively studied in the context of galaxy clusters \citep[see][]{mcnamara07, fabian12}. This is especially relevant at Cosmic Noon, where heightened accretion and star-forming activity may result in a larger fraction of molecular gas excited to warm (T$\sim 500$ K) or hot (T $\sim 1000-3000$ K) phases. The warm molecular gas is typically traced via rotational transitions of the H$_2$ molecule in rest-frame mid-infrared spectra \citep[e.g.,][]{armus23, hernandez23, Esparza-Arredondo25}. In this paper, we focus on the hot molecular gas phase at traced by the ro-vibrational transitions of H$_2$ in rest-frame near-infrared spectra, particularly the 1-0 S(1) and 1-0 S(0) transitions at 2.12 $\mu$m and 2.22 $\mu$m, respectively \citep[e.g.,][]{black87, hollenbach89, maloney96, elmegreen89, rosario19, riffel21, riffel23}.

\begin{figure*}
\centering
\includegraphics[width=0.4\textwidth]{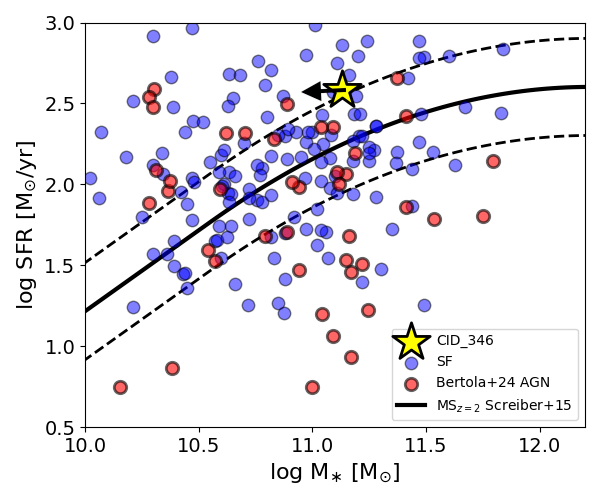}
\includegraphics[width=0.4\textwidth]{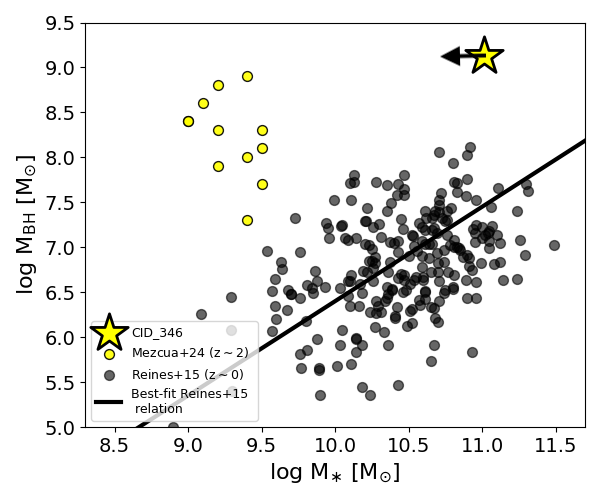}
\includegraphics[width=0.4\textwidth]{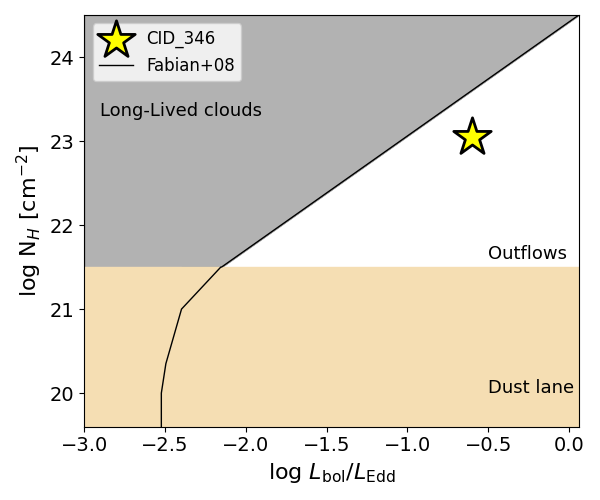}
\includegraphics[width=0.4\textwidth]{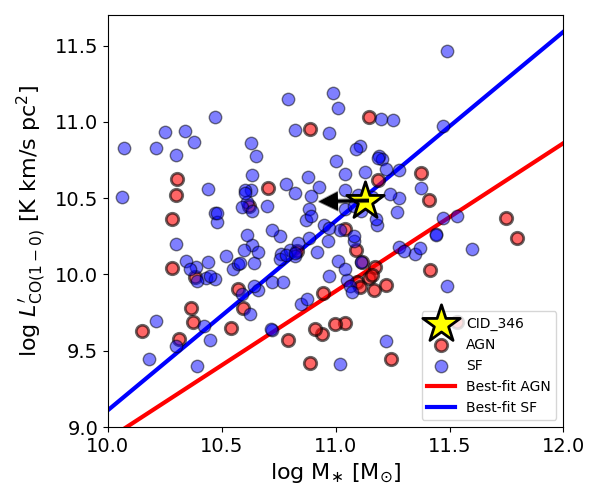}
\caption{{\it Top left panel} shows the location of cid\_346, at the top end of the star forming main sequence (at $z\sim 2$). The solid line shows the main sequence relation from \citet{schreiber15} with 0.3 dex error margins shown as dashed lines. The red and blue data points show the z$\sim$2 AGN and star forming galaxies, respectively \citep[e.g.,][]{tacconi18, boogaard20, birkin21, bertola24}. The {\it top right panel} shows the location of cid\_346 in the $M_{BH}-M_{\ast}$ plane. cid\_346 is clearly above the $M_{BH}-M_{\ast}$ relation for local galaxies (solid black line and black circles from \citet{reines15}), similar to recent results from cosmic noon \citep[yellow circles from][]{mezcua24} and galaxies at the Epoch of Re-ionisation.  The {\it bottom left panel} shows the location of cid\_346 in the outflowing region of the $N_{\rm H}$ versus $\lambda_{\rm Edd}$ parameter space \citep[see][]{fabian08}. The {\it bottom right panel} shows cid\_346 in the L$^{\prime}_{\rm CO(1-0)}$ versus M$_{\ast}$ plane, with the best-fit lines from \citet{bertola24}. The colour scheme in the bottom right panel is the same as in the top left panel. For further details, please refer to Sect. \ref{sect2}.}
\label{fig:cid_346_props}
\end{figure*}

At low-redshift, these ro-vibrational and rotational transitions have been extensively observed in a variety of galaxies, including galaxies that host AGN-driven outflows \citep[e.g.,][]{davies09, storchi-bergmann09, friedrich10, dasyra11, rupke13, ramos-almeida17, rosario19, riffel21, armus23, hernandez23, Esparza-Arredondo25, ulivi25}. The picture arising from these low-redshift studies is that near-infrared and mid-infrared H$_{2}$ lines are strongly enhanced in cool gas shocked by AGN-driven outflows. Therefore, at high redshift, especially at cosmic noon, where several AGN and star forming galaxies have been shown to host outflows, the warm and hot molecular gas are expected to be prevalent. Up until the launch of JWST, it was not possible to trace such warm and hot molecular transitions at cosmic noon. However, with the unique spectroscopic capabilities of the Medium resolution Spectrometer (MRS) of the Mid-Infrared Instrument (MIRI) on board JWST, it is now possible to trace the rest-frame near-infrared transitions out to cosmic noon and beyond. 

\begin{figure}
\centering
\includegraphics[width=0.95\linewidth]{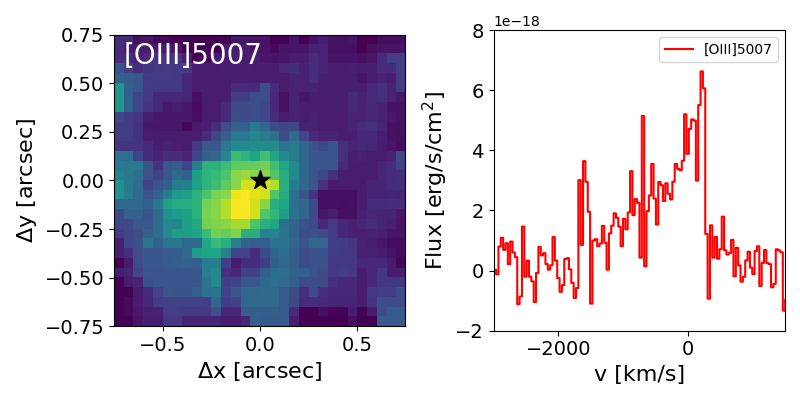}
\includegraphics[width=0.95\linewidth]{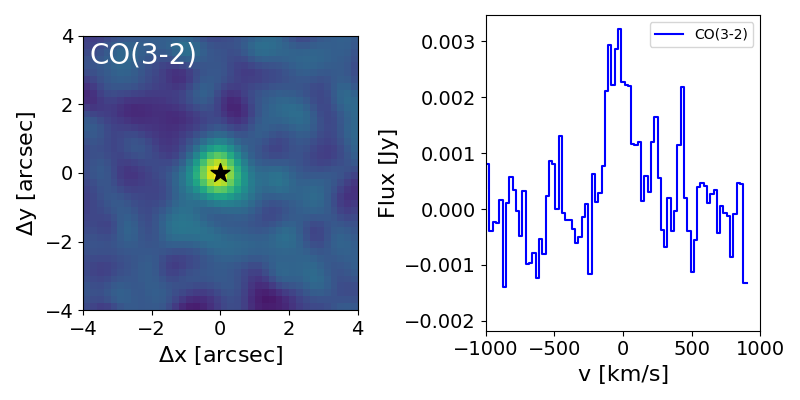}
\caption{The {\it top panels} show the \oiii~ flux map (left) and the integrated spectrum covering the entire region of the \oiii ~emission (right). The \oiii ~is extended towards the SE of the AGN location (marked by black star) and its spectrum shows a prominent blue-wing, indicative of ionised outflows. The {\it bottom panels} show the ALMA CO(3-2) flux map (left) and the corresponding integrated spectrum (right). Further details about the ionised and molecular gas properties are described in Sect. \ref{sect2}.}
\label{fig:cid_346_oiii_co}
\end{figure}

In this paper, we present the first-ever detection of the hot molecular gas transition in an AGN host galaxy at $z\sim 2.2$, traced via rest-frame near-infrared ro-vibrational H$_{2}$ transitions. The selected target is a hard X-ray AGN source, which hosts kpc-scale ionised gas outflows, traced using the \oiii ~line and is detected in CO(3-2) transition via ALMA and ACA observations. The paper is arranged as follows: Section \ref{sect2} summarises the basic properties of the target, followed by a description of JWST/MRS mid-infrared spectroscopic and JWST/NIRCam near-infrared imaging observations and data reduction in section \ref{sect3}. The analysis of the mid-infrared spectra and NIRCam images and the corresponding results are presented in section \ref{sect4}. This is followed by a discussion and concluding remarks in Sects. \ref{sect5} and \ref{sect6}, respectively. 

The following cosmological parameters are adopted throughout this paper: $H_{0}$ = 70 km s$^{-1}$, $\Omega_{\rm M}$ = 0.3 and $\Omega_{\Lambda}$ = 0.7. Unless specified, North is up and East is to left in the maps presented. 

\section{Target Description} \label{sect2}

The target, cid\_346 (RA (J2000) = 09:59:43.4, DEC (J2000) = $+$02:07:07.4, $z=2.219$: redshift based on CO(3-2)), is part of the SINFONI Survey for Unveiling the Physics and Effect of Radiative feedback \citep[SUPER; e.g.,][]{circosta18, mainieri21} and is located within the COSMOS field \citep[e.g.,][]{scoville07, civano16}. Publicly available multi-wavelength photometric and spectroscopic data have enabled precise determination of the AGN and host galaxy properties through Spectral Energy Distribution (SED) modeling \citep[see][]{circosta18}. In this section, we outline the key characteristics of this target, also summarised in Fig. \ref{fig:cid_346_props}, and refer to \citet{circosta18, circosta21}, \citet{vietri20}, and \citet{bertola24}\footnote{The SED fitting results for cid\_346 were updated in \citet{bertola24} using the latest releases of the photometric catalogs for the COSMOS field \citep{jin18, weaver22}.} for further details. 

cid\_346 is classified as a Type-1 AGN, as indicated by the presence of broad lines in its rest-frame ultraviolet (UV) and optical spectra. The black hole mass, estimated from the broad H$\beta$ line, is log $M_{\rm BH}$ = 9.15$\pm$0.30 M$_{\odot}$ \citep[see][]{vietri20}. Stellar mass and SFR, derived through SED modelling, are log M$_{\ast}$ $<$ 11.13 M$_{\odot}$ and 382$\pm$37 M$_{\odot}$ yr$^{-1}$, respectively (see \citet{bertola24}). These values position cid\_346 at the upper boundary of the star-forming main sequence in the SFR-M$_{\ast}$ plane (Fig. \ref{fig:cid_346_props}, top left panel). Its black hole mass is significantly higher than expected based on the local M$_{BH}$-M$_{\ast}$ relation, suggesting that the supermassive black hole in cid\_346 may be over-massive relative to its host galaxy (Fig. \ref{fig:cid_346_props}, top right panel). This phenomenon, where over-massive black holes are observed, has been reported in recent studies in galaxies at the Cosmic Noon to the Epoch of Re-ionisation \citep[e.g.,][]{ubler23, mezcua24}. The position of cid\_346 in these scaling relations, in conjunction with mid-infrared data and comparisons to other AGN host galaxies, will be discussed in a forthcoming paper. 

The bolometric luminosity of the AGN in cid\_346 is log $L_{\rm bol} = 46.66 \pm 0.02$ erg s$^{-1}$. Based on the black hole mass reported earlier, the Eddington ratio of the AGN, $\lambda_{\rm Edd}$ (where $\lambda_{\rm Edd} = L_{\rm bol}/L_{\rm Edd}$ and $L_{\rm Edd} = 1.3 \times 10^{38} \times M_{\rm BH}/M_{\odot}$ erg s$^{-1}$), is 0.25. Additionally, X-ray spectral analysis reveals that cid\_346 hosts a highly obscured AGN, with an X-ray column density log $N_{\rm H} = 23.05^{+0.17}_{-0.19}$ cm$^{-2}$. We place these values in the $N_{\rm H}$ versus $\lambda_{\rm Edd}$ plane, within the context of empirical models of outflows in dusty ISM from \citet{fabian08} in Fig. \ref{fig:cid_346_props}. Below log $N_{\rm H}=21.5$ cm$^{-2}$ (horizontal line in lower left panel in Fig. \ref{fig:cid_346_props}), dust lanes from the host galaxy could contaminate the column density values. The empirical continuous line in the figure divides the region where Eddington ratio experienced by the cloud is below the Eddington limit (grey shaded region) or above the Eddington limit (unshaded region), the latter defined as the "forbidden region" for long-lived clouds or "blowout region" \citep[see also][]{kakkad16, ricci17b}. cid\_346 falls within the unshaded region of the $N_{\rm H}$ versus $\lambda_{\rm Edd}$ plot, and is therefore expected to experience an outflow.

cid\_346 has been the focus of several observational campaigns using various ground-based telescopes to trace extended ionised and cold molecular gas. A brief overview of these observations is provided below, with key results relevant to this paper summarised in Figure \ref{fig:cid_346_oiii_co}.
\begin{itemize}
\item SINFONI near-infrared spatially resolved IFS of cid\_346 in the H-band reveals ionised gas outflow traced by \oiii, extending up to $\sim$4 kpc in the southeast direction from the nucleus. This extended outflow is further confirmed through PSF-subtraction techniques \citep[e.g.,][]{kakkad20}. The outflow velocity exceeds 1000 km s$^{-1}$, indicating that the AGN is driving the outflow. Based on the outflow model described in \citet{kakkad20}, the ionised gas mass outflow rate, $\dot{M}_{\rm out}$, is estimated to be between 0.6–15 M$_{\odot}$ yr$^{-1}$. This is significantly lower than the star formation rate of $\sim$380 M$_{\odot}$ yr$^{-1}$, suggesting that there is still sufficient gas in the host galaxy to support ongoing star formation despite the fast outflow. 

SINFONI K-band IFS observations revealed that H$\alpha$ emission is associated with the Narrow Line Region (NLR) outflow, as it extends in the same direction as the ionised gas, with FWHM $>$600 km s$^{-1}$. A resolved BPT map of the extended gas, combined with the \oiii~results from the H-band, indicates that this emission is most likely ionised by the AGN, based on emission line ratio diagnostic diagrams commonly used for the local universe \citep[e.g.,][]{kewley01, kauffmann03}.

\item cid\_346 was also observed with low spatial resolution ($\sim$1.1 arcsec) ALMA Band-3 to observe CO(3-2) transition to trace the total cold molecular gas content \citep[see][]{circosta21}. The CO(3-2) luminosity of cid\_346, log $L^{\prime}_{\rm CO(3-2)}$, was estimated to be 10.25 K km s$^{-1}$ pc$^{2}$. Assuming an excitation correction ($r_{31} = L^{\prime}_{\rm CO(3-2)}/L^{\prime}_{\rm CO(1-0)}$) of 0.59 and a CO to molecular gas conversion factor, $\alpha_{\rm CO}$, of 3.6 M$_{\odot}$/(K km s$^{-1}$ pc$^{2}$), the total cold molecular gas content in cid\_346 is log $M_{\rm mol} = 11.10\pm0.05$ M$_{\odot}$. The CO-based cold gas to M$_{\ast}$ ratio is consistent with the loci of star forming galaxies in the $L^{\prime}_{\rm CO(1-0)}$ versus M$_{\ast}$ plot shown in the bottom right panel of Figure \ref{fig:cid_346_props}. The background blue and red points are obtained from a compilation of molecular gas studies in star forming and AGN host galaxies in \citet{bertola24} \citep[see also][]{tacconi18, boogaard20, birkin21}, which uses the following excitation corrections: $r_{41} = 0.37, r_{31} = 0.59, r_{21} = 0.68$ \citep[e.g.,][]{kirkpatrick19}. This plot confirms that cid\_346 contains an abundance of molecular gas for sustained star formation or for feeding the central supermassive black hole.
\item Follow-up Atacama Compact Array (ACA) observations of cid\_346 were obtained in order to map the large scale CO emission out to the circum-Galactic Scales \citep[CGM, $>$100 kpc, see][]{cicone21}. Although the extension of CO gas out to CGM scales was later disputed in \citet{jones23}, these observations reveal that cid\_346 hosts extended molecular gas out to at least tens of kiloparsec scales. At the time of publishing these results, no evidence of merging companions were detected close to cid\_346. Therefore, one of the possible explanations put forward was that the outflows from the AGN might relocate the molecular gas out to CGM scales. 
\end{itemize}

Figure \ref{fig:cid_346_oiii_co} presents a summary of the [OIII]$\lambda$5007 and CO(3-2) images and spectra from SINFONI and ALMA observations, respectively. The observations summarised above came from ground-based observatories, which were limited in spatial resolution and depth. However, they consistently pointed to the presence of extended gas in the ionised and molecular phases. The ionised gas phase is believed to be part of the extended NLR of cid\_346. However, with the launch of JWST, it has become possible to trace rest-frame optical and near-infrared emissions with unprecedented spatial resolution and depth, surpassing even the best Adaptive Optics capabilities of ground-based telescopes. 

\section{Observations and data reduction} \label{sect3}

\subsection{JWST MIRI/MRS observations and data reduction} \label{sect3.1}

cid\_346 was observed with MIRI/MRS aboard JWST as a part of Cycle-1 GO programme 2177 (PI: Mainieri). Target acquisition was not required during the observations as the pointing accuracy of 0.1 arcsec is sufficient to locate the target in the MRS field-of-view. Furthermore, we employed a 4-point dither pattern, which enhanced the spatial sampling and spatial resolution in the final drizzled data cube \citep[see][]{law23}. We also took dedicated background exposures for background subtraction, as the contribution of infrared background is substantial in Channels 3 and 4. 

The MRS observations were carried out on 20 November 2023 in the LONG band. The LONG band simultaneously covers the wavelength ranges 6.53--7.65 $\mu$m, 10.02--11.70 $\mu$m, 15.41--17.98 $\mu$m and 24.19--27.90 $\mu$m in Channels 1, 2, 3 and 4 respectively. At the redshift of cid\_346, these correspond to rest-frame wavelength ranges of 2.02--2.37 $\mu$m, 3.11--3.63 $\mu$m, 4.78--5.58 $\mu$m and 7.51--8.66 $\mu$m in the respective channels. The respective resolving power, R, in each channel is 3100--3610, 2860--3300, 1980--2790 and 1630--1330. The observations employed a SLOW readout mode in 38 groups per integration, 3 integrations per exposure and 12 exposures. As a result, the final total exposure time on-source was 11084s ($\sim$3.07 hours).

We mostly follow the standard MRS data reduction procedure, as set by the pipeline \citep[e.g.,][]{labiano16} with additional customised steps and fine-tuning parameters to improve data quality. We reduced the data using the JWST pipeline version 1.17.1. The data reduction steps are briefly summarised here. The pipeline runs in three stages. The first stage of the MRS pipeline performs the detector-level corrections. We kept most of the default pipeline parameters unchanged, except we set the \texttt{find\_showers} keyword to \texttt{TRUE}. This step identifies and corrects the Cosmic Ray (CR) showers. To improve the identification of warm pixels, which may have been missed in the stage-1 data reduction, we derived the median of all slope images of each detector and applied a classical sigma clipping algorithm. The newly identified bad and warm pixels were masked in the bad pixel mask as \texttt{DO\_NOT\_USE}. The output of the first stage is a set of corrected count rate files, which are then processed with the second stage of the pipeline. The second stage performs instrument calibrations such as wavelength calibration, flat-field correction, flux calibration and removes fringes. For the second stage, the default pipeline parameters were unchanged, except the \texttt{skip\_residual\_fringe} parameter, which was set to \texttt{FALSE} (Default is \texttt{TRUE}). This parameter corrects the fringing patterns in the individual channels. The background subtraction was performed using dedicated background exposures. The second stage of the MRS pipeline outputs a set of fully calibrated individual exposures. These fully calibrated exposures are then finally fed into the third stage of the pipeline, where these calibrated exposures across different wavelength bands and channels are combined into a final drizzled data cube. The parameters for the third stage were also kept unchanged, except the \texttt{skip\_outlier\_detection} was set to \texttt{FALSE}, with the outlier threshold kept at 99\%, which removes any significant remaining outliers in the data when combining different exposures. 

The final products from the three stages of the pipeline consist of a set of four data cubes, corresponding to the four channels of the LONG configuration. The field-of-view in channels 1, 2, 3 and 4 are $\approx 3.2 \times 3.7$ arcsec$^{2}$, $4.0\times 4.8$ arcsec$^{2}$, $5.2\times 6.2$ arcsec$^{2}$ and $6.6\times 7.7$ arcsec$^{2}$, respectively. The corresponding spatial resolution  of the channels are $\approx$0.18, 0.28, 0.39 and 0.64 arcsec, respectively. 

\subsection{NIRCam and ACS imaging from COSMOS} \label{sect3.2}
cid\_346 was also observed as a part of the COSMOS-Web survey, a JWST Cycle-1 treasury programme (GO-1727, PI: Kartaltepe \& Casey, see \citet{casey23} for an overview). The COSMOS-Web survey covered 0.54 $\mathrm{deg}^2$ of the COSMOS survey \citep[e.g.,][]{scoville07} with four filters of the NIRCam instrument \citep[F115W, F150W, F277W, F444W][]{rieke23}. In this paper, we use the NIRCam imaging of cid\_346 to compare with the H$_{2}$ maps in the MRS data in Sect. \ref{sect4}.

The NIRCam observations were taken in April 2023, and the data were processed using the JWST Calibration Pipeline version 1.14.0 \footnote{\url{https://github.com/spacetelescope/jwst}} \citep[see][]{Bushouse2024_jwst1140} and the Calibration Reference Data System (CRDS) version 1223. In addition to NIRCam imaging, we also used the publicly-available archival HST/ACS F814W imaging \citep{koekemoer07}. The NIRCam and ACS images have a pixel scale of 0.03 arcsec/pixel.

\section{Analysis and results} \label{sect4}
In this section, we present the detection of the hot molecular gas via ro-vibrational transitions in the MRS spectra. Furthermore, we will present evidence for the presence of companions close to cid\_346, with the help of ancillary NIRCam imaging data from COSMOS-Web. Finally, we will compare the MRS and NIRCam results with ground-based SINFONI and ALMA observations of this target. 

\subsection{Detection of hot molecular gas} \label{sect4.1}

% Location of Satellite A = 09:59:43.336 +02:07:06.60; Location of Sat B = 09:59:43.280, +02:07:06.090

\begin{figure}
\centering
\includegraphics[width=0.5\textwidth]{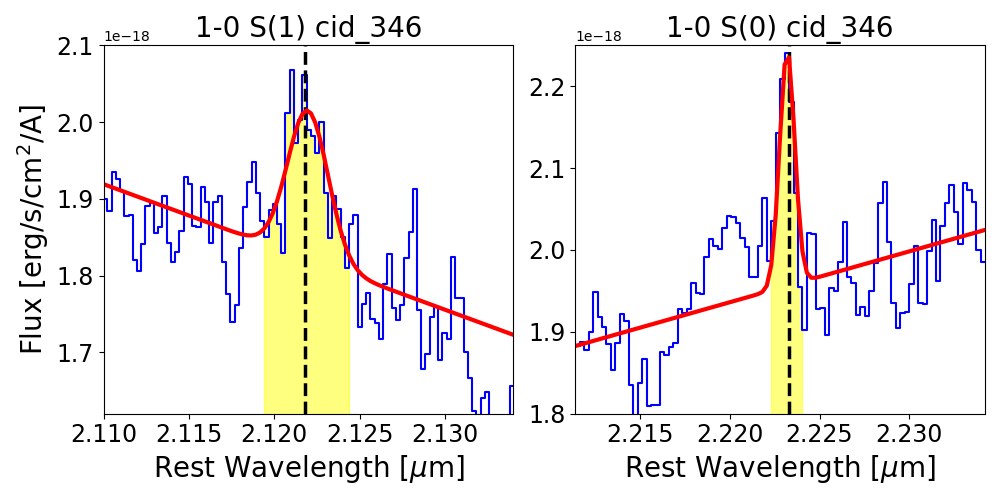}
\caption{1-0 S(1) and 1-0 S(0) detections in the central AGN, cid\_346. The blue curve shows the data extracted from the MRS cubes, the red curve shows the single Gaussian models to the emission lines, along with local continuum, and the yellow shaded region shows the spectral window containing $\sim$95\% ($\pm 2\sigma$) of the emission line flux. The black dashed line shows the expected location of each line, based on the redshift of cid\_346.}
\label{fig:warmH2_cid_346}
\end{figure}

\begin{table*}
\centering
\begin{tabular}{lcccccc}
\hline
Transition & Expected wavelength & Actual wavelength & Red/Blue-shift & FWHM & Flux & SNR\\
& $\mu$m & $\mu$m & (+/-)km s$^{-1}$ & km s$^{-1}$ & $10^{-18}$erg s$^{-1}$ cm$^{-2}$\\
(a) & (b) & (c) & (d) & (e) & (f) & (g)\\
\hline\hline
{\bf cid\_346} \\
1-0 S(1) & 2.1218 & 2.1220$\pm$0.0002 & 23$\pm$22  & 458$\pm$60 & 5.7$\pm$0.8 & 4.1\\
1-0 S(0) & 2.2233 & 2.2232$\pm$0.0001 & $-$8$\pm$8  & 130$\pm$20 & 2.6$\pm$0.4 & 4.4\\
\hline
{\bf C1} \\
1-0 S(1) & 2.1218 & 2.1209$\pm$0.0003 & $-$130$\pm$40 & 500$\pm$125 & 2.5$\pm$0.6 & 2.4\\
\hline
{\bf C2} \\
1-0 S(1) & 2.1218 &  2.1223$\pm$0.0004 & 67$\pm$24 & 127$\pm$61 & 1.6$\pm$0.5 & 4.1\\
\hline\hline
\end{tabular}
\caption{Detected H$_{2}$ ro-vibrational transitions tracing the hot molecular gas in the central AGN host, cid\_346 and locations C1 and C2 (see Fig. \ref{fig:warmH2_maps}). (a) reports the ro-vibrational H$_{2}$ transition, (b) reports the expected rest-frame wavelength of the transition, based on the redshift quoted in \citet{circosta21}, (c) reports the actual wavelength of the transition, based on the centroid location of the Gaussian fit to the line, (d) shows the shift of the measured line location with respect to the expected location. $+$ means the line is red-shifted and $-$ means the line is blue-shifted, (e) reports the line width (FWHM), (f) shows the line flux and (g) reports the signal-to-noise (SNR) of the detected line. Parameters reported in columns (c), (d) and (e) are from single Gaussian fits, as described in Sect. \ref{sect4.1}. The errors represent 1$\sigma$ deviation from the quoted values.}
\label{tab:warmH2_fit_results}
\end{table*}

Before the launch of JWST, sensitive mid-infrared spectrographs were unavailable, making MRS spectroscopy a unique opportunity to study the rest-frame near-infrared spectra of galaxies, particularly at Cosmic Noon. Among the various lines within near-infrared and mid-infrared wavelengths, the ro-vibrational and rotational transitions of hydrogen molecules are used as tracers of molecular gas at temperatures from a few hundred to several thousand Kelvin. In this paper, we focus on the hot H$_{2}$ phase in the rest-frame near-infrared wavelengths, that could be excited by fluorescence soft-UV photons in both star forming and AGN-ionised regions \citep[e.g.,][]{black87} or via thermal processes from X-ray or shock heating \citep[e.g.,][]{maloney96, hollenbach89}. 

We are particularly interested in the 1-0 S(1) transition at rest-wavelength 2.12 $\mu$m (we use the notation 1-0 S(1) and H$_{2}$ 2.12 $\mu$m interchangeably), commonly used to trace hot molecular gas in nearby galaxies and falls in channel-1 LONG sub-band. Therefore, we first extract the channel-1 LONG spectra from the AGN location, identified using the peak of the continuum emission, using a circular aperture of radius $\sim$0.4 arcsec. This aperture size approximately covers the spatial extent of the central host galaxy (we detect extended gas components, which are discussed later). In order to trace possible residual instrument response feature in the MRS spectra, we also selected three object-free regions, randomly selected in the data cube, from where spectra were extracted and compared with the target spectrum. This procedure was also used to confirm the line detections described here. A comparison between the target spectrum and spectrum from one of the object-free locations is shown in Fig. \ref{fig:H2_BKG}. 

Following this comparison, we confirm detections of the 1-0 S(1) and 1-0 S(0) transitions in cid\_346 at $\sim$4$\sigma$ levels each. A zoom-in of the extracted MRS spectra around these transitions is shown in Fig. \ref{fig:warmH2_cid_346} in blue. We model these H$_{2}$ lines using a single Gaussian function, along with a linear continuum that models the local continuum within the spectral window centred on the individual transitions. We do not impose any constraints while fitting these Gaussian profiles to the emission lines, with the exception that the line peak intensities are positive. The Gaussian models to the emission lines are shown in red in Fig. \ref{fig:warmH2_cid_346} and the yellow regions mark the flux containing 95\% of the emission line flux ($\pm$2$\sigma$). To estimate the errors in our measurements, we generated 100 mock spectra by adding Gaussian random noise to the modelled spectra, using the standard deviation of line-free regions of the spectra. The fitting procedure described earlier was then applied to each of these mock cubes without imposing any parameter constraints. The associated error is given by the standard deviation of these 100 measurements. The parameters corresponding to the Gaussian models, namely the line centroid, width (FWHM) and the fluxes are reported in Table \ref{tab:warmH2_fit_results}. 

At the sensitivity of the MIRI/MRS data, we do not detect the presence of extended wings in the spectral profiles of the H$_{2}$ transitions. This is confirmed by the fact that it was not necessary to add a second Gaussian component during the line fitting. The width (FWHM) of the 1-0 S(1) and 1-0 S(0) are 458$\pm$60 and 130$\pm$20 km s$^{-1}$, respectively. Given a resolving power of $\sim$100 km s$^{-1}$ at the location of these lines, the transitions are well resolved. In previous studies, a cut of 600 km s$^{-1}$ has usually been employed to distinguish between gas outflows in the ionised phase that are driven by an AGN or not \citep[e.g.,][]{kakkad20, wylezalek20, tozzi24}. This value is based on the maximum velocity dispersion one expects in a purely rotating disk in a star forming galaxy. The transitions reported in the MRS spectra do not cross this threshold. It is also possible that the observations did not have sufficient depth to detect the presence of faint wings beneath the emission line profiles. However, we note that the bulk of the molecular gas, whether based on CO or ro-vibrational transition, may be dominated by gas in the host galaxy disk \citep[see also][]{barfety25}. In addition, simulations by \citet{ward24} showed that in the presence of a clumpy ISM, the outflow can cover a broad range of velocities, with a considerable fraction of mass moving below 100 km s$^{-1}$. Therefore, the cut of 600 km s$^{-1}$ used in case of the ionised gas, largely coming from the NLR, may not be applicable here. The H$_{2}$ line widths reported in Table \ref{tab:warmH2_fit_results} are significantly narrower than that of \oiii-based ionised gas from ground-based IFU observations, which show a FWHM of $>$1500 km s$^{-1}$. 

\begin{figure}
\centering
\includegraphics[width=0.9\linewidth]{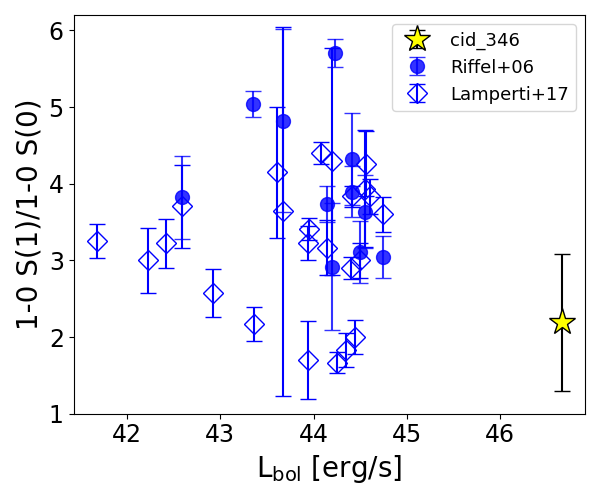}
\caption{1-0 S(1) to 1-0 S(0) line ratio in cid\_346 (yellow star) and low-redshift Seyfert galaxies (blue circles) as a function of the bolometric luminosity. The error bars represent 1$\sigma$ uncertainties in the respective line ratios.}
\label{fig:h2_ratios}
\end{figure}

We find the flux ratio, 1-0 S(1) to 1-0 S(0), to be $\sim$2.2$\pm$0.7. The S(1):S(0) flux ratio can be used to determine the temperature of the hot molecular gas and a comparison between samples spanning a wide range of redshift and luminosity can provide clues into the impact of the presence of AGN on the molecular gas temperatures, discussed later in this section. We compare the 1-0 S(1) to 1-0 S(0) flux ratio in cid\_346 with measurements in low-redshift AGN host galaxies from \citet{riffel06} and \citet{lamperti17} in Fig. \ref{fig:h2_ratios}. \citet{riffel06} consists of a compilation of near-infrared spectra of 47 $z<0.5$ AGN and \citet{lamperti17} consists of a compilation of 102 $z<0.075$ X-ray selected AGN from BAT AGN Spectroscopic Survey \citep[BASS, e.g.,][]{koss17}. Figure \ref{fig:h2_ratios} shows only those low-redshift targets from these studies which have detections and/or upper limits on both 1-0 S(1) and 1-0 S(0) transitions. Both studies consist of measurements from type-1 and type-2 AGN. Our measured flux ratio for cid\_346 is within the range of typical flux ratios of the two lines observed in the comparison low-redshift AGN host galaxies. The line ratio also appears to be at the lower end of the range seen in low-redshift galaxies. However, this is based on a single galaxy at the high luminosity end and that we need more number statistics at these higher luminosities, especially for $L_{\rm bol} > 10^{45}$ erg s$^{-1}$, to confirm the trend. Further details on the possible origin of the H$_{2}$ emission in cid\_346 are discussed in Sect. \ref{sect5}.

We can use the 1-0 S(1) and 1-0 S(0) transitions to constrain the hot gas temperature using the following equation \citep[e.g., see][]{wilman05}:

\begin{equation}
{\rm log} \left(\frac{F_{\rm i}\lambda_{\rm i}}{A_{\rm i}g_{\rm i}}
\right) = {\rm constant} - \frac{T_{\rm i}}{T_{\rm exc}}
\label{eq:h2_exc}
\end{equation}

\noindent
In Eq. \ref{eq:h2_exc}, \(F_{\rm i}\) represents the flux of the \(i\)th H$_{2}$ transition, \(\lambda_{\rm i}\) is the associated wavelength, \(A_{\rm i}\) denotes the spontaneous emission coefficient, \(g_{\rm i}\) is the statistical weight of the upper level of the transition, \(T_{\rm i}\) refers to the energy of the level expressed as a temperature, and \(T_{\rm exc}\) is the kinetic temperature assuming the H$_{2}$ is in equilibrium. Using the equation above for the 1-0 S(1) and 1-0 S(0) transitions, we derive a hot gas temperature of 1100$^{+465}_{-245}$ K for the central source cid\_346.

We can also estimate the mass of the hot molecular gas using 1-0 S(1) transition at 2.12 $\mu$m, adopting the standard prescriptions used for low-redshift galaxies in the literature \citet{storchi-bergmann09, riffel23}. The choice of 1-0 S(1) transition is motivated by the fact that it is one of the strongest ro-vibrational transitions. As a result, there are several low-redshift observations of this line for which the calculation is well-established and would make it ideal for comparing our mass measurements. Specifically, we use the following equation to determine the mass of the hot H$_{2}$ gas:

\begin{equation}
M_{\rm H_{2}} = \frac{2~m_{\rm p}F_{2.12}~4\pi D_{\rm L}^{2}}{f_{\rm \nu=1,J=3}A_{\rm S(1)}h\nu}
\label{eq:h2_mass}
\end{equation}

\noindent
where $m_{p}$ is the proton mass, $F_{2.12}$ is the flux of the 1-0 S(1) line, $D_{L}$ is the luminosity distance, $f$ is the population fraction and $A_{S(1)}$ is the transition probability. The population fraction is obtained directly from quantum mechanical principles and we obtain transition probabilities from \citet{turner77} (see also \citet{wolniewicz98} and \citet{roueff19}). For the vibrational temperature of T$\sim$1100$^{+465}_{-245}$ K, $f_{\rm \nu=1,J=3}$ takes values between $1-7.5 \times 10^{-3}$. Using the flux of the 1-0 S(1) line reported in Table \ref{tab:warmH2_fit_results} into Eq. \ref{eq:h2_mass}, we obtain a hot H$_{2}$ gas mass of $5.0^{+7.0}_{-3.5} \times 10^{5}$ M$_{\odot}$ in cid\_346.

We can now compare the mass of the molecular gas in cold phase ($<$100 K) to the hot phase ($\sim$1000 K -- 3000 K). The total amount of CO-based cold gas in the cid\_346 system is $1.2 \times 10^{11} M_{\odot}$ \citep[see][]{circosta21}. The ratio between the CO-based cold molecular gas mass and the H$_{2}$ 2.12 $\mu$m-based hot molecular gas mass is $\approx 10^{5}-10^{6}$, i.e., the CO-based cold gas mass is nearly six orders of magnitude larger than the H$_{2}$ 2.12 $\mu$m-based hot gas mass, similar to the values seen in some of the low-redshift galaxies \citep[e.g.,][]{Pereira-Santaella16, emonts17}. We further discuss the comparison of cold-to-hot gas mass ratios with low redshift galaxies in Sect. \ref{sect5}.

\subsection{Extended hot molecular gas} \label{sect4.2}

\begin{figure}
\centering
\includegraphics[width=0.45\textwidth]{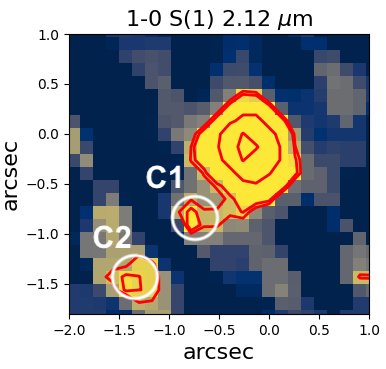}\\
\caption{1-0 S(1) channel map obtained by collapsing the channels centred on 2.12 $\mu$m emission. The red contours show emission above 3$\sigma$. The H$_{2}$ maps indicate extended clumps of hot gas around cid\_346. In addition to the central host, reference spectra were extracted from the locations marked C1 and C2, where the extended emission is detected in most transitions, $\sim$0.8 and $\sim$1.5 arcsec away from cid\_346, respectively.}
\label{fig:warmH2_maps}
\end{figure}

\begin{figure}
\centering
\includegraphics[width=0.5\textwidth]{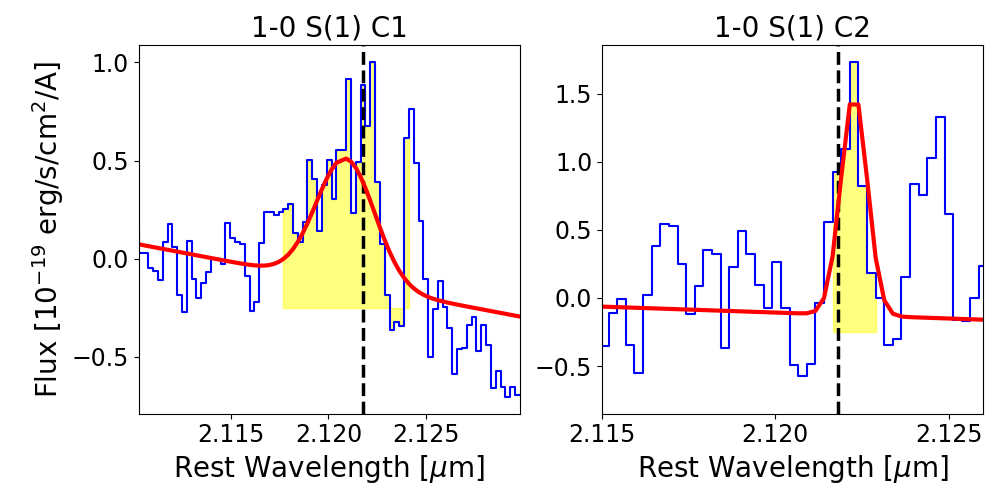}
\caption{1-0 S(1) detections in extended regions: the left panel shows spectrum from C1 and the right panel shows spectrum from C2 (see Fig. \ref{fig:warmH2_maps}).}
\label{fig:warmH2_extended}
\end{figure}

\begin{figure}
\centering
\includegraphics[width=0.5\textwidth]{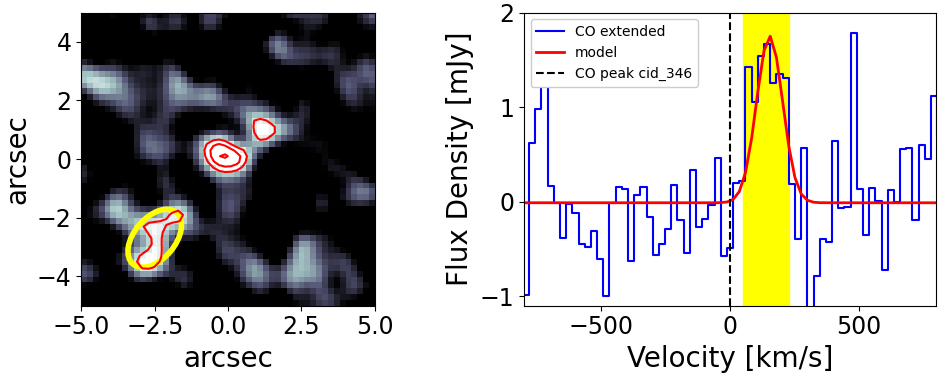}
\caption{The left panel shows a diffuse extended CO(3-2) emission towards the SE of cid\_346 at (0,0) arcsec. The red contours show the CO emission at levels 3$\sigma$, 4$\sigma$ and 5$\sigma$. The spectrum extracted from the yellow circle, covering the extended CO emission, is shown in the right panel. The blue curve in the right panel shows the data and the red curve shows the Gaussian model to the emission line. The dashed black line shows the location of the CO(3-2) peak emission from cid\_346 (see also \citet{circosta21}), suggesting that the extended CO(3-2) emission is red-shifted with respect to cid\_346.}
\label{fig:extended_co}
\end{figure}

In this section, we present channel maps of the brighter 1-0 S(1) line and search for the presence of extended hot molecular gas emission. Figure \ref{fig:warmH2_maps} shows the hot molecular gas map, obtained by collapsing the channels containing 1-0 S(1) emission. The hot molecular gas appears extended beyond the MRS PSF and the observed spatial extent is observed in approximately the same direction as the \oiii-based ionised gas reported in \citet{kakkad20}. In particular, we identified two locations, which we name C1 and C2 approximately 1.0 ($\sim$8.2 kpc) and 2.0 ($\sim$16.4 kpc) arcsec from the AGN location, respectively, (also marked in Fig. \ref{fig:warmH2_maps}), where the hot molecular gas emission is prominent. Within the current scope of this paper, we do not attempt to decompose emission contributions from the QSO and host galaxy separately \citep[e.g.,][]{chen24} and will be implemented in future studies utilising this MIRI-MRS data. This would also facilitate a direct comparison with existing ground-based ancillary near-infrared IFU and sub-mm data.

We extract MRS spectra from C1 and C2 using circular apertures centred on the respective locations, with a radius of 0.25 arcsec, approximately corresponding to the PSF of the MRS at these wavelengths \footnote{https://jwst-docs.stsci.edu/jwst-mid-infrared-instrument/miri-operations/miri-dithering/miri-mrs-psf-and-dithering}. The size of the apertures were manually chosen to roughly correspond to the extent of the emission in each spatial component. We detect 1-0 S(1) emission in C1 and C2 at 2.4$\sigma$ and $\sim$4$\sigma$, respectively (see Figs. \ref{fig:warmH2_maps} and \ref{fig:warmH2_extended} and Table \ref{tab:warmH2_fit_results}). The 1-0 S(0) component remained undetected in the extended regions.

Having detected the extended hot molecular gas in the MIRI/MRS data in the same direction as the extended ionised gas in VLT/SINFONI data, we re-analysed the archival ALMA CO(3-2) data of cid\_346 (Project code: 2016.1.00798.S) to search for extended cold molecular gas component in the SE direction of the AGN. The details of the ALMA programme from which we obtained the data are available in \citet{circosta21}. The left panel in figure \ref{fig:extended_co} shows the detection of diffuse and extended CO(3-2) emission, located $>$2.0 arcsec from cid\_346 at $\sim$3$\sigma$ significance level. We extracted a spectrum from a circular aperture covering the extended CO emission, shown as a yellow circle in Fig. \ref{fig:extended_co}. The extracted CO spectrum is shown in the right panel of Fig. \ref{fig:extended_co}. The peak location of the extended CO component is red-shifted by $\sim$120 km s$^{-1}$. The width of the extended CO component is 120 km s$^{-1}$, compared to a width of 201 km s$^{-1}$ observed in cid\_346.

\begin{figure}
\includegraphics[scale=0.3]{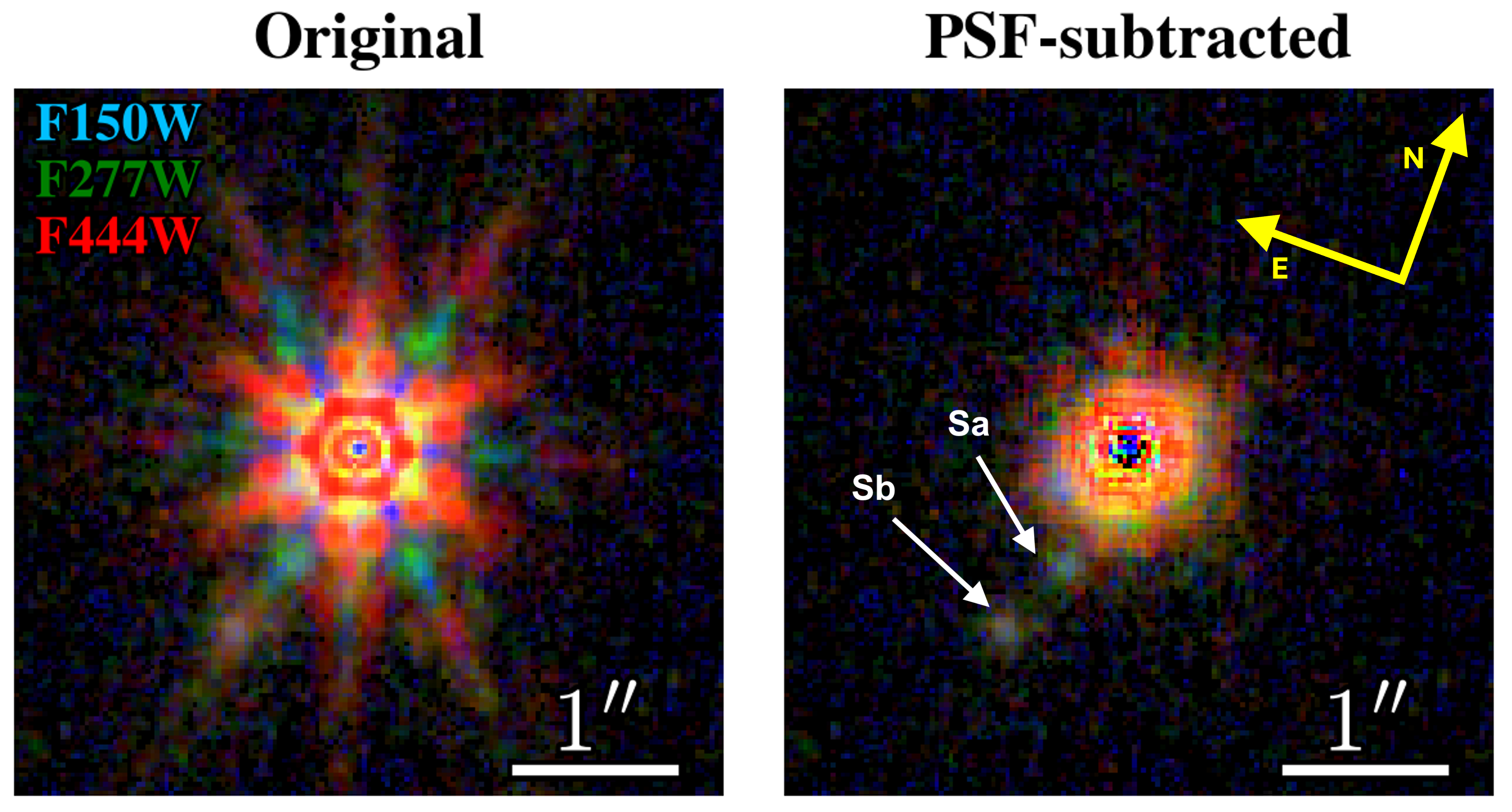}
\caption{The left panel shows the combined multi-colour image from the raw NIRCam image filters F150W, F277W and F44W and the right panel shows the corresponding PSF-subtracted image where the two companions, Sa and Sb, are clearly visible.}
\label{fig:cid346_nircam_multicolour}
\end{figure}

\begin{figure}
\centering
\includegraphics[width=0.45\textwidth]{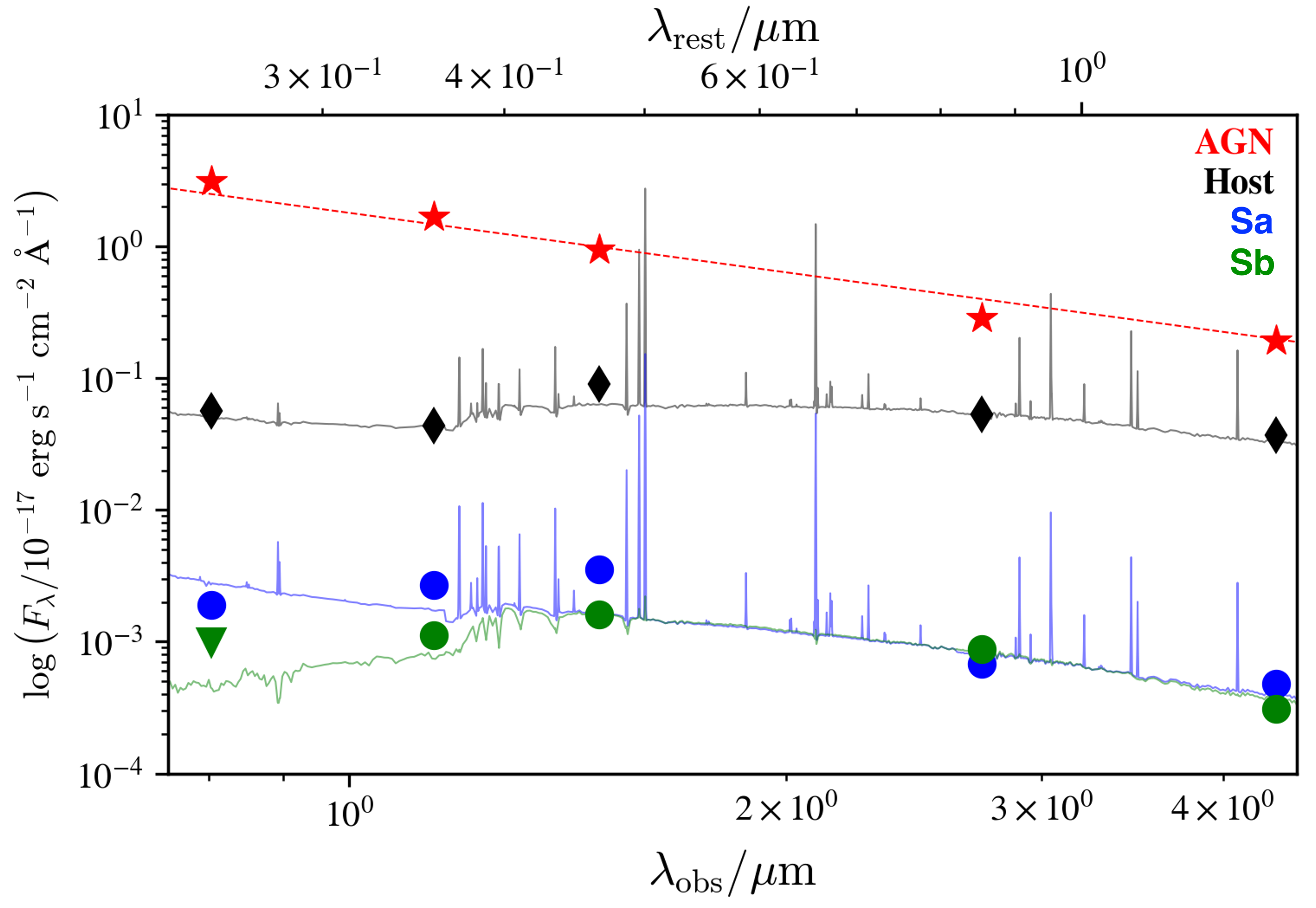}
\caption{The SED fitting results of cid\_346, Sa and Sb. The red data points and the corresponding curve shows the AGN component of cid\_346, the black colours corresponds to the host galaxy of cid\_346, the blue colours to Sa and green to Sb. }
\label{fig:cid346_sed}
\end{figure}

Following the procedure described in Sect. \ref{sect4.1}, we calculate the  gas mass in the extended regions, C1 and C2, to be $1.5^{+2.3}_{-1.1} \times 10^{5}$ M$_{\odot}$ and 1.6$^{+2.3}_{-1.1}$ $\times$10$^{5}$ M$_{\odot}$, respectively. Along with the central host, cid\_346, the total hot molecular gas in cid\_346 system is $\sim 8.4\times 10^{5}$ M$_{\odot}$. Most of the hot molecular gas ($\sim$60\%) is concentrated in the central AGN itself, while the rest of the hot gas is in the extended regions. Similarly, the total amount of CO-based cold gas in the cid\_346 system is $1.5 \times 10^{11} M_{\odot}$, out of which $\sim$80\% is from cid\_346 \citep[1.2$\times10^{11}$ M$_{\odot}$, see][]{circosta21}, while the rest 20\% is from the extended component. Even after accounting for these extended components, cold molecular gas dominates the mass budget. 

\begin{table*}
\centering
\caption{Morphological and photometric parameters of cid\_346 system estimated from the four-component image modelling. See Sect. \ref{sect4.3} for further details.}\label{tab:psf_pars}
\begin{threeparttable}[t]
\begin{tabular}{cccccccc}
\hline
Component & $n$\tnote{a} & $r_{\rm e}/\mathrm{arcsec}$\tnote{a} & $m_\mathrm{F814W}$\tnote{b} & $m_\mathrm{F115W}$\tnote{b} & $m_\mathrm{F150W}$\tnote{b} & $m_\mathrm{F277W}$\tnote{b} & $m_\mathrm{F444W}$\tnote{b}\\
\hline\hline
AGN PSF & - & - & $19.3\pm 0.1$ & $19.2\pm 0.1$ & $19.2\pm 0.1$ & $19.3\pm 0.03$ & $18.7\pm 0.02$ \\
AGN host & $0.8\pm0.2$ & $0.25\pm0.01$ & $23.6\pm 0.6$ & $23.1\pm 0.5$ & $21.7\pm 0.3$ & $21.13\pm 0.07$ & $20.5\pm 0.1$ \\
Sa & $2.5\pm0.6$ & $0.15\pm0.05$ & $26.9\pm 0.4$ & $26.3\pm 0.4$ & $25.3\pm 0.2$ & $25.91\pm 0.07$ & $25.1\pm 0.2$ \\
Sb & $1.3\pm0.2$ & $0.10\pm0.01$ & $<$27.8 & $26.8\pm 0.4$ & $26.0\pm 0.2$ & $25.87\pm 0.06$ & $25.9\pm 0.1$ \\
\hline
\end{tabular}
\begin{tablenotes}
\item[a] $n$ and $r_e$ are the values measured for the F277W image.
\item[b] AB magnitude \citep{OkeAB}.
\end{tablenotes}
\end{threeparttable}
\end{table*}

\subsection{Companions around cid\_346 with NIRCam imaging} \label{sect4.3}
As described in Sect. \ref{sect3.2}, cid\_346 was also covered as a part of the COSMOS-Web imaging campaign. Given the extended ionised, hot and cold molecular gas, we explored the imaging data to search for the presence of possible companions around cid\_346. 

JWST/NIRCam images are contaminated by PSF-diffracted spikes. Therefore, we cannot visually identify the presence of extended component in the host galaxy or possible presence of companions in each of the filters (F150W, F277W and F444W) of NIRcam raw images (see left panel in Fig. \ref{fig:cid346_nircam_multicolour}). Therefore, we perform PSF-subtraction based on the methods outlined in \citet{tanaka24} and briefly mentioned here. NIRCam PSFs were constructed using  \texttt{PSFEx} \citep[see][]{bertin11} that empirically models PSF based on stars detected from applying \texttt{SeXtractor} software package \citep[see][]{bertin96}. We fit the images from each NIRCam band with a composite model consisting of a single PSF component and additional S\'{e}rsic components. The number of additional components required was decided based on the lowest Bayesian Information Criterion \citep[BIC, see][]{schwarz78}. 

We required a total of three S\'{e}rsic components to fit the cid\_346 system. The first S\'{e}rsic component was required for the host galaxy of the AGN itself. In addition, we fit two other S\'{e}rsic components, which we refer as Sa and Sb, corresponding to two companions which are detected towards the SE of the AGN location. We estimate the error of the fitted parameters using $\chi^2_\nu$-based weighting of each single PSF, following the approach of \cite{Ding2020}. Table \ref{tab:psf_pars} reports the values of the parameters $n$, $r_{\rm e}$, flux, and S/N of each of the components estimated from the above four-component fitting. 

The left panel of Fig. \ref{fig:cid346_nircam_multicolour} shows the raw NIRCam multi-colour image combining the filters F150W, F277W and F444W and the right panel shows the PSF-subtracted multi-colour image where the two components are detected. Based on the F277W results, the centre positions of Sa and Sb are (RA (J2000) = 09:59:43.4, DEC (J2000) = $+$02:07:06.5) and (RA (J2000) = 09:59:43.4, DEC (J2000) = $+$02:07:06.0), respectively, which are offset by $\sim$0.8 arcsec ($\sim$6.6 kpc) and $\sim$1.4 arcsec ($\sim$11.6 kpc) from the host galaxy. The raw images of the individual NIRCam images and the PSF subtracted images are moved to the appendix (see Figure \ref{fig:cid346_nircam_allfilters}). 

\begin{figure*}
\centering
\includegraphics[scale=0.5]{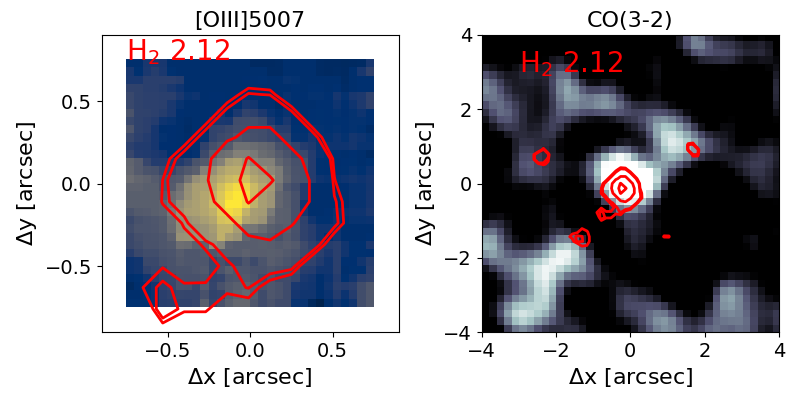}
\includegraphics[scale=0.5]{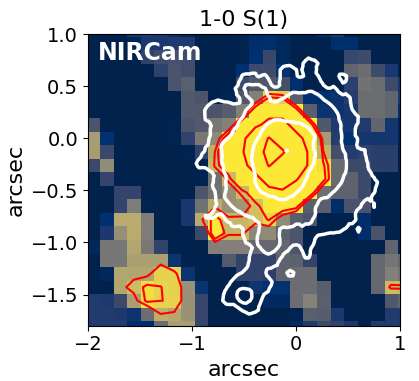}
\caption{The background image in the {\it left panel} shows the ground-based \oiii ~flux map from SINFONI observations \citep{kakkad20}, the {\it middle panel} shows the extended CO(3-2) channel map from ALMA observations from Fig. \ref{fig:extended_co} and the {\it right panel} shows the 1-0 S(1) map (also shown in Fig. \ref{fig:warmH2_maps}). The red contours in the all the panels show the H$_{2}$ 1-0 S(1) emission at 2.12 $\mu$m and the white contours in the right panel show the NIRCam PSF-subtracted contours.}
\label{fig:NC_H2_overlay}
\end{figure*}

The indications of companions (or satellite galaxies), along with extended multi-phase gas in cid\_346 suggests that this galaxy may be in a minor merger. Although we will later show that the components, C1 and C2, in H$_{2}$ emission discussed in Sect. \ref{sect4.2} do not directly align with the Sa and Sb detected in NIRCam images (right panel in Fig. \ref{fig:NC_H2_overlay}), the presence of companions may explain the extended gas reservoir seen in cid\_346 \citep[see also][]{cicone21}. We also estimate the properties of each companion using Spectral Energy Distribution (SED) fitting, following the methods in \citet{tanaka24}. We briefly describe the procedure here. We fit the photometric fluxes of each component using \texttt{CIGALE-v2022.1} SED fitting library \citep[see][]{boquien19, yang22}. As the extended hot molecular gas is detected in a similar direction as the Sa and Sb components, we assume that Sa and Sb are at the same redshift as cid\_346. We model the stellar populations using a single stellar population model \citep[by][]{bruzual03}, assuming a Chabrier initial mass function \citep[IMF;][]{chabrier03} with a stellar mass ($M_{\ast}$) range between 0.1--100 M$_{\odot}$. We assume a delayed-$\tau$ model for a star-formation history (see equation 7 in \citet{tanaka24}). We model the nebular emission using the \texttt{nebular} module and dust attenuation using the \texttt{dustatt\_modified\_starburst} module, the latter assuming a modified \cite{calzetti00} law. To limit the models within physical solutions, we set the upper limit of $t_{\rm age}$ to $0.95 t_H$, where $t_H$ indicates the cosmic age at each redshift. Stellar metallicity, $Z_{\ast}$ and the ionisation parameter, $U$, are fixed at Solar metallicity and $\log U=-2$, respectively.

Fig. \ref{fig:cid346_sed} compares the best-fit model SED with the photometry of each of the components. We estimate the spectral slope of the AGN component, $\alpha$ ($F_\lambda \propto \lambda^\alpha$), to be $\sim1.5$, which is typical for nearby AGNs \citep[e.g.,][]{VandenBerk2001}. The consistency of the slope with that of nearby AGN validates the decomposition of the AGN and PSF components presented earlier. The components Sa and Sb, on the other hand, exhibit different SED shapes, which may suggest different SFR and dust properties of each of these galaxies. From the SED fitting results, Sa has a SFR = 1.6$\pm$0.9 M$_{\odot}$ yr$^{-1}$, while Sb has SFR = 0.5$\pm$0.6 M$_{\odot}$ yr$^{-1}$, suggesting Sa may have higher recent star formation activity than Sb. The stellar mass of the AGN host, log $M_{\ast}$, from the SED fits is 10.9$\pm$0.3 M$_{\odot}$. This value is consistent with the literature results of cid\_346 in \citet{tanaka24} and similar to the SED fitting presented in \citet{circosta18} and \citet{bertola24}. The stellar mass of the two companions, Sa and Sb, on the other hand, are log $M_{\ast}$ = 8.9$\pm$0.3 M$_{\odot}$ and 8.9$\pm$0.2 M$_{\odot}$, respectively. Although this suggests that the companion galaxies are 100 times less massive than the AGN host itself, the SED fitting presented have large uncertainties and the analysis presented here only indicates the possible presence of a minor merger in cid\_346 system. We further discuss the role of possible mergers in driving the AGN activity in Sect. \ref{sect5}.

\section{Discussion} \label{sect5}

% To discuss/cite this paper: https://ui.adsabs.harvard.edu/abs/2024arXiv241006960R/abstract

% Regarding H2 to BrY ratio, cite this paper: https://ui.adsabs.harvard.edu/abs/2021MNRAS.503.5161R/abstract

% Regarding H2 2.12 to 2.2233 ratio, cite: Riffel+06 paper

We presented new JWST rest-frame near-infrared spectra and NIRCam imaging of cid\_346, a luminous AGN at cosmic noon epoch. Figure \ref{fig:NC_H2_overlay} shows a summary of the extended emission from the multi-wavelength spatially-resolved data presented in this paper and in the literature: \oiii-based ionised gas maps from VLT/SINFONI, CO(3-2)-based cold molecular gas maps from ALMA, H$_{2}$ 2.12 $\mu$m-based hot molecular gas from MIRI/MRS and PSF-subtracted NIRCam F277W image. All the gas components i.e., ionised gas, cold and hot molecular gas are extended approximately in the same spatial direction (SE of the AGN). There are differences in the exact locations and extension of the different gas components. VLT/SINFONI observations only probed the inner 1.5 arcsec of the host galaxy (or 0.75 arcsec $\equiv$ 6 kpc from the AGN location). However, based on the cold and hot molecular gas observations, it is possible that the ionised gas might be extended beyond 6 kpc from the AGN. The peak of the extended CO(3-2) emission is seen nearly at the edge of the hot molecular gas clump C2. This suggests a multi-phase ionisation or gas excitation by the AGN, starting with the warm ionised gas phase ($\sim$10000-20000 K), followed by hot molecular ($\sim$1000 K) and cold molecular ($<$100 K) gas, as one moves away from the AGN location. 

Over the last few years, several observations of the cold molecular gas in mass-matched AGN and non-AGN host galaxies at Cosmic Noon have revealed a systematic difference between the CO-based cold molecular gas content in the two classes of galaxies. In these studies, the AGN host galaxies seem to have lower CO-based cold gas mass compared to non-AGN hosts \citep[see][]{kakkad17, brusa18, circosta21,bertola24}. A few observations at low-redshift have revealed that molecular gas exists in warm or hot gas phases, where the CO emission is absent \citep[e.g.,][]{rosario19, davies24}. Our JWST/MRS and ALMA observations of cid\_346 were mainly motivated to evaluate what fraction of molecular gas may exists in the warm or hot molecular gas phases at cosmic noon, compared to the cold molecular gas phase. The observations presented in this paper may be one of the first detections of H$_{2} ~2.12 \mu$m-based hot molecular gas in a galaxy at Cosmic Noon. Here we present some possibilities on the mechanisms in cid\_346 that could excite the detected ro-vibrational transitions. The H$_{2}$ ro-vibrational transitions in the near-infrared spectra could be excited by non-thermal processes such as fluorescent excitation due to absorption of UV photons \citep[e.g.,][]{black87} or by thermal processes such as excitation by X-ray heating \citep[e.g.,][]{maloney96} or by shocks \citep[e.g.,][]{hollenbach89}. Several line ratios and correlations can be used to distinguish between the different excitation mechanisms. For example, the line ratio: H$_{2} ~\lambda2.2477/2.1218$ can distinguish between thermal ($\sim$0.1) and fluorescent ($\sim$0.5) excitation mechanisms \citep[e.g.,][]{mouri94, reunanen02, rodriguez-ardila05, riffel08, storchi-bergmann09}. We do not detect H$_{2}$ 2.2477 $\mu$m emission in the MIRI-MRS data of cid\_346. However, we calculate an upper limit on the flux ratio H$_{2} ~\lambda2.2477/2.1218$ of 0.3, which suggests that the observed H$_{2} ~2.12 ~\mu$m is likely excited by thermal processes. We can also estimate the 2.12 $\mu$m flux that can be generated by a given X-ray incident radiation, following \citet{maloney96} models. We follow the methodology adopted in \citet{riffel08} \citep[see also][]{zuther07, storchi-bergmann09}, calculating the ionisation parameter, $\xi_{\rm eff}$, using incident hard X-ray flux at distances of $\approx$100-200 pc from the central source. We calculate the expected emergent H$_{2} ~2.12 ~\mu$m flux in the range $\sim 10^{-18}-10^{-17}$ erg s$^{-1}$ cm$^{-2}$, for a $\sim$0.4 arcsec aperture, the same as the one used to extract the MRS spectrum from the central source in cid\_346. Comparing this emergent flux with the observed flux reported in Table \ref{tab:warmH2_fit_results}, it is likely that the X-ray heating from the AGN can account for most of the observed H$_{2}$ 2.12 $\mu$m flux.

% Shock heating: Kristensen et al. 2023: A&A, 675, A86

\begin{figure}
\centering
\includegraphics[width=0.8\linewidth]{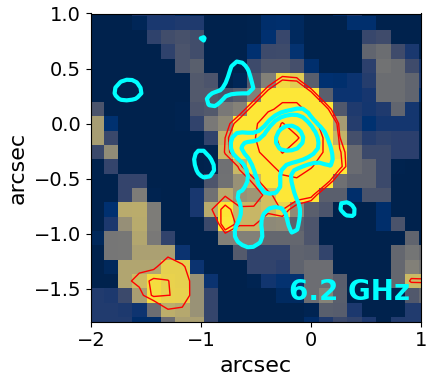}
\caption{Background image shows the 2.12 $\mu$m flux map with 6.2 GHz radio contours from VLA A-array (Ilha et al.). }
\label{fig:radio_h2_overlay}
\end{figure}

Fig. \ref{fig:radio_h2_overlay} shows the 2.12 $\mu$m emission with 6.2 GHz radio contours obtained from VLA A-array observations overlaid. We matched the peak of the radio emission and the peak of H2 2.12 um emission for the overlay shown in Fig. \ref{fig:radio_h2_overlay}. The extended radio structure around this source bending approximately at the location of C1 suggests that the C1 component/clump may have been shock-excited due to the radio jet. Even in the central source, the radio jets may contribute to the H$_{2}$ emission. Details of the JVLA observations will be presented in detail in a forthcoming paper (Ihla et al.). Therefore, in addition to, or alternative to the X-ray radiation from AGN accretion disc, radio jets might deplete CO-based molecular gas while enhancing the emission from hot molecular gas. Coupled with the fact that NIRCam imaging observations suggest cid\_346 is in an interacting system, the emerging picture from these observations is that the hot molecular gas forms a part of the larger extended gas reservoir of this merging system of galaxies, that is shock-heated due to the presence of radio jets and is also being ionised by the central AGN.

Using the observed flux and temperature of the H$_{2} ~2.12 ~\mu$m line, we find the total mass of the hot molecular gas phase, calculated using the 2.12 $\mu$m emission extracted from cid\_346, C1 and C2 (Fig. \ref{fig:warmH2_maps}), of $8\times 10^{5}$ M$_{\odot}$. The ratio of hot-to-cold molecular gas in cid\_346 is $10^{-6}-10^{-5}$. Using a combination of VLT/SINFONI H$_{2}$ 2.12$\mu$m and ALMA CO-based observations, \citet{emonts17} and \citet{Pereira-Santaella16} reported a hot-to-cold molecular gas mass ratio of $\sim 6-7 \times10^{-5}$ in low-redshift ($z\sim 0.01$) Luminous Infrared Galaxies (LIRGs) hosting either an AGN or starburst. Although the statistics are low, comparing our results from cid\_346, with that of the local LIRGs, there does not appear to be a significant evolution in the hot-to-cold molecular gas masses with redshift, at least until cosmic Noon. On the other hand, \citet{dale05} reported lower hot-to-cold gas mass ratios of $\sim 10^{-7}-10^{-5}$ in a heterogenous sample of nearby galaxies, but predominantly normal star forming galaxies. The lower hot-to-cold gas mass ratios in non-AGN galaxies may suggest that instead of a redshift evolution in the ratio, it may be the AGN luminosity that drives the hot molecular gas masses to higher values. However, more statistics are required at Cosmic Noon that includes both star forming and AGN host galaxies to isolate the redshift evolution and AGN impact on molecular gas. {Focusing only on the hot-to-cold gas ratio of cid\_346, } our results from JWST/MRS spectroscopy, therefore, suggest that at cosmic noon, the bulk of the molecular gas still exists in the cold gas phase. When comparing the mass of total molecular gas in AGN versus non-AGN host galaxies after taking into account the hot molecular gas mass, we would still end up with a systematic difference in the molecular gas content between the two classes of galaxies. On the other hand, as mentioned earlier in this section, the findings in this paper reveals that the impact of the AGN on molecular gas may be observed on localised spatial scales. Future JWST/MRS observations of both AGN and star forming galaxies at cosmic noon are required to expand our measurements of the relative amount of hot and cold molecular gas phases and the detectability of the warm gas phase in a general population of galaxies at that epoch. An MRS$+$ALMA study of a statistical sample of galaxies will also reveal if there are systematic differences in the hot molecular gas properties between mass-matched AGN and star forming galaxies, as is revealed with ALMA for the cold gas phase \citep[e.g.,][]{kakkad17, circosta21, bertola24}.

PSF-subtracted NIRCam images of cid\_346 from COSMOS-Web survey reveals the presence of at least two companions roughly towards the south east of the AGN location. This may explain the presence of extended multi-phase gas reservoir in this system, although the exact locations of the continua may not correspond to the locations of the spatial peaks or clumps of the extended gas, as shown in the right panel of Fig. \ref{fig:NC_H2_overlay}. The indications of satellite galaxies may suggest a merger-driven growth of both the supermassive black hole as well as star formation in the host galaxy of cid\_346. At the same time, the accretion appears to drive multi-phase outflows, clearly seen in the \oiii-based ionised gas and tentative indications in the hot molecular gas. However, we make a few observations about the outflows in different gas phases: First: In ground-based IFU data, it is possible that red-shifted and blue-shifted components of the emission lines (from the inflow or outflow or simply due to turbulence in the merger) got blended with each other, which may artificially broadened emission lines. Second: The absence of wings in molecular gas tracers compared to ionised gas could be due to the fact that molecular gas might be dominated by emission from the disk of the host galaxy, while the \oiii ~emission may be dominated by emission from the NLR. At low-redshift, for example in the interacting system Arp 220 \citep[see][]{perna24}, wings representing outflowing gas has been observed in Pa-$\alpha$ or iron lines (\ion{Fe}{ii}), but such broad line profiles are not detected in the H$_{2}$ transitions, similar the observations here \citep[see also][]{speranza22}.  On the other hand, limited SNR or the limited spectral resolution in the H$_{2}$ spectra may also limit the detection rates of outflow wings underneath the lines. Therefore, future deeper MRS observations of this source are required to reveal potential wings in the individual emission line profiles and  diffuse hot molecular gas emission around the cid\_346 system.

\section{Conclusions and Final remarks} \label{sect6}

\noindent
We analysed MIRI/MRS spectra and NIRCam images of an AGN host galaxy at Cosmic noon, cid\_346. Previous ground-based spatially-resolved observational campaigns of cid\_346 already showed the presence of extended outflowing ionised gas with a large reservoir of cold molecular gas. The primary aim of the MRS spectroscopy was to search for the hot molecular gas phase using the ro-vibrational transitions in the rest-frame near-infrared spectra such as the 1-0 S(1) transition at 2.12 $\mu$m. The following summarises the findings presented in this paper:

\begin{itemize}
\item We detect ro-vibrational transition of the H$_{2}$ molecule, particularly the 1-0 S(1) and 1-0 S(0) lines. The molecular lines are likely excited by X-ray heating from the AGN. We find an excitation temperature of 1100$^{+465}_{-245}$ K and a total hot molecular gas of $\sim 8.0\times 10^{5} M_{\odot}$. The hot molecular gas mass is nearly five to six orders of magnitude less than ALMA CO-based cold gas mass, similar to hot-to-cold molecular gas mass ratios observed at low-redshift. This suggests that cold molecular gas dominates the bulk of the molecular gas budget in this galaxy and that the hot-to-cold molecular gas ratio does not appear to change with redshift.
\item We find extended hot molecular gas towards the SE of the AGN location up to a distance of 2.0 arcsec ($\sim$16 kpc). The extended emission is roughly in the same direction as the extended ionised gas observed in ground-based IFU observations. A re-analysis of ALMA data also revealed extended CO(3-2) emission at $\sim3\sigma$ level. Nearly 60\% of hot molecular gas is in the central AGN host, cid\_346 and the rest in extended regions. In the case of cold molecular gas, 80\% is in the central AGN host and 20\% in the extended regions. Comparing the H$_{2}$ 2.12 $\mu$m emission with VLA A-array 6.2 GHz radio emission suggests that the extended ro-vibrational transitions may be excited by the radio jets.
\item NIRCam imaging reveals the presence of two satellite galaxies near cid\_346. Photometric and SED fitting analysis suggests that the stellar mass of these satellites is $\sim$100 times lower than that of the central AGN host galaxy, cid\_346. Collectively, the cid\_346 system is abundant with multi-phase gas, extended to tens of kiloparsec scales from the central AGN location. 
\end{itemize}

The results presented in this paper showcases the capability of JWST mid-infrared spectrographs to detect hot molecular gas in galaxies at Cosmic noon. Future surveys targeting a statistical sample of both AGN and star forming galaxies will be essential to derive a multi-phase molecular gas picture at an epoch where feedback from AGN and/or star formation is expected to play a pivotal role. Additionally, deep resolved observations with both JWST and ALMA are required to understand the origins of the extended multi-phase molecular gas. Furthermore, comparison with zoom-in simulations on the impact of radiation on molecular gas composition will be crucial to infer the impact of AGN on preventative mode feedback by heating molecular gas to higher temperatures. 

\section*{Acknowledgements}
We thank the anonymous referee for their constructive comments and suggestions.
This work used the CANDIDE computer system at the Institut d’Astrophysique de Paris, which was funded through grants from the PNCG, CNES, DIM-ACAV, and the Cosmic Dawn Center and maintained by S. Rouberol. Numerical computations were in part carried out on the iDark cluster, Kavli IPMU. Kavli IPMU is supported by World Premier International Research Center Initiative (WPI), MEXT, Japan. 
TT is supported by JSPS KAKENHI Grant Number 25KJ0750 and the Forefront Physics and Mathematics Program to Drive Transformation (FoPM), a World-leading Innovative Graduate Study (WINGS) Program at the University of Tokyo.
EB and GC acknowledge financial support from INAF under the Large Grant 2022 ``The metal circle: a new sharp view of the baryon cycle up to Cosmic Dawn with the latest generation IFU facilities''. 
GSI acknowledges financial support from the Fundação de Amparo à Pesquisa do Estado de São Paulo (FAPESP), under Projects 2022/11799-9 and 2024/02487-9. S.C acknowledges support by European Union’s HE ERC Starting Grant No. 101040227 - WINGS. 
MP acknowledges grant PID2021-127718NB-I00 funded by the Spanish Ministry of Science and Innovation/State Agency of Research (MICIN/AEI/ 10.13039/501100011033), and the grant RYC2023-044853-I, funded by  MICIU/AEI/10.13039/501100011033 and European Social Fund Plus (FSE+). 
RAR acknowledges the support from Conselho Nacional de Desenvolvimento Cient\'ifico e Tecnol\'ogico (CNPq; Proj. 303450/2022-3, 403398/2023-1, \& 441722/2023-7), Funda\c c\~ao de Amparo \`a pesquisa do Estado do Rio Grande do Sul (FAPERGS; Proj. 21/2551-0002018-0), and Coordena\c c\~ao de Aperfei\c coamento de Pessoal de N\'ivel Superior (CAPES;  Proj. 88887.894973/2023-00). 
GT acknowledges financial support from the European Research Council (ERC) Advanced Grant under the European Union’s Horizon Europe research and innovation programme (grant agreement AdG GALPHYS, No. 101055023). 
IL acknowledges support from PRIN-MUR project “PROMETEUS”  financed by the European Union -  Next Generation EU, Mission 4 Component 1 CUP B53D23004750006. 
CMH acknowledges funding from an United Kingdom Research and Innovation grant (code: MR/V022830/1).
This paper makes use of the following ALMA data: ADS/JAO.ALMA\#2016.1.00798.S. ALMA is a partnership of ESO (representing its member states), NSF (USA) and NINS (Japan), together with NRC (Canada), NSC and ASIAA (Taiwan), and KASI (Republic of Korea), in cooperation with the Republic of Chile. The Joint ALMA Observatory is operated by ESO, AUI/NRAO and NAOJ.

%%%%%%%%%%%%%%%%%%%%%%%%%%%%%%%%%%%%%%%%%%%%%%%%%%
\section*{Data Availability}
The data used in this paper is available in the MAST portal of STScI and in ESO and ALMA archives. 

%%%%%%%%%%%%%%%%%%%% REFERENCES %%%%%%%%%%%%%%%%%%

% The best way to enter references is to use BibTeX:

\bibliographystyle{mnras}
\bibliography{reference} % if your bibtex file is called example.bib

% Alternatively you could enter them by hand, like this:
% This method is tedious and prone to error if you have lots of references
%\begin{thebibliography}{99}
%\bibitem[\protect\citeauthoryear{Author}{2012}]{Author2012}
%Author A.~N., 2013, Journal of Improbable Astronomy, 1, 1
%\bibitem[\protect\citeauthoryear{Others}{2013}]{Others2013}
%Others S., 2012, Journal of Interesting Stuff, 17, 198
%\end{thebibliography}

%%%%%%%%%%%%%%%%%%%%%%%%%%%%%%%%%%%%%%%%%%%%%%%%%%

%%%%%%%%%%%%%%%%% APPENDICES %%%%%%%%%%%%%%%%%%%%%

\appendix

\section{Extra Material}

\begin{figure*}
\centering
\includegraphics[width=0.65\textwidth]{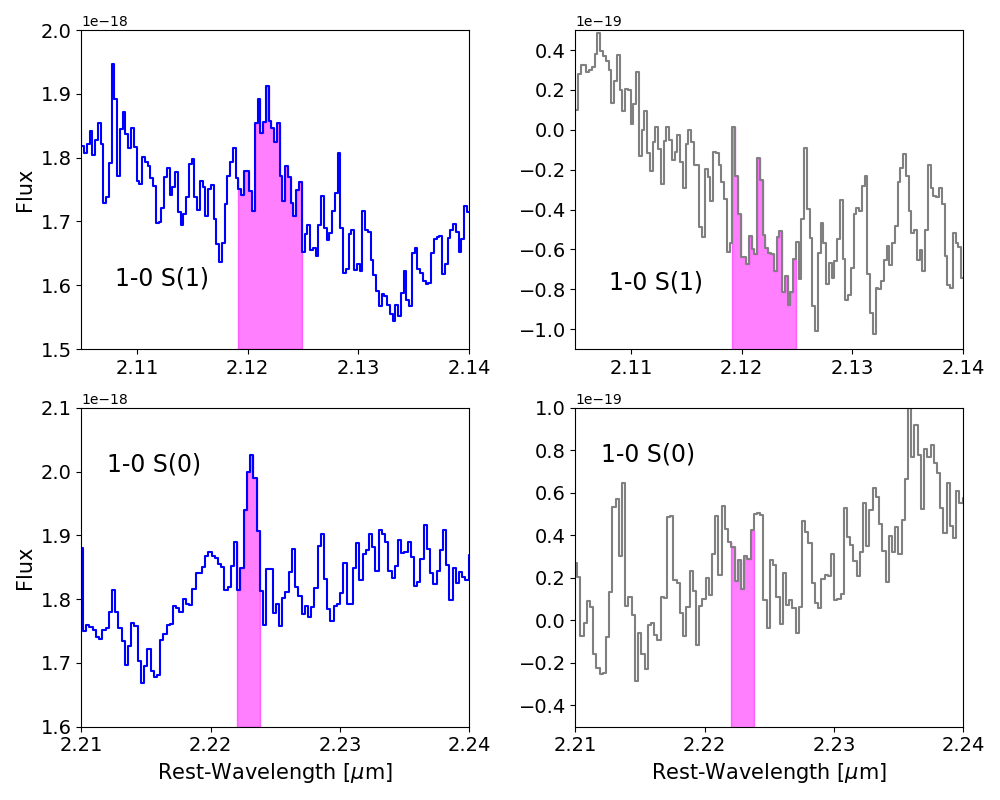}
\caption{Comparison of H$_{2}$ detections in cid\_346 integrated spectra (blue curves in the left panels) with spectra extracted from object-free regions (grey curve in the right panel). The magenta shaded regions mark the location of the 1-0 S(1) and 1-0 S(0) lines in the top and bottom panels, respectively.}
\label{fig:H2_BKG}
\end{figure*}

\begin{figure*}
\includegraphics[scale=0.5]{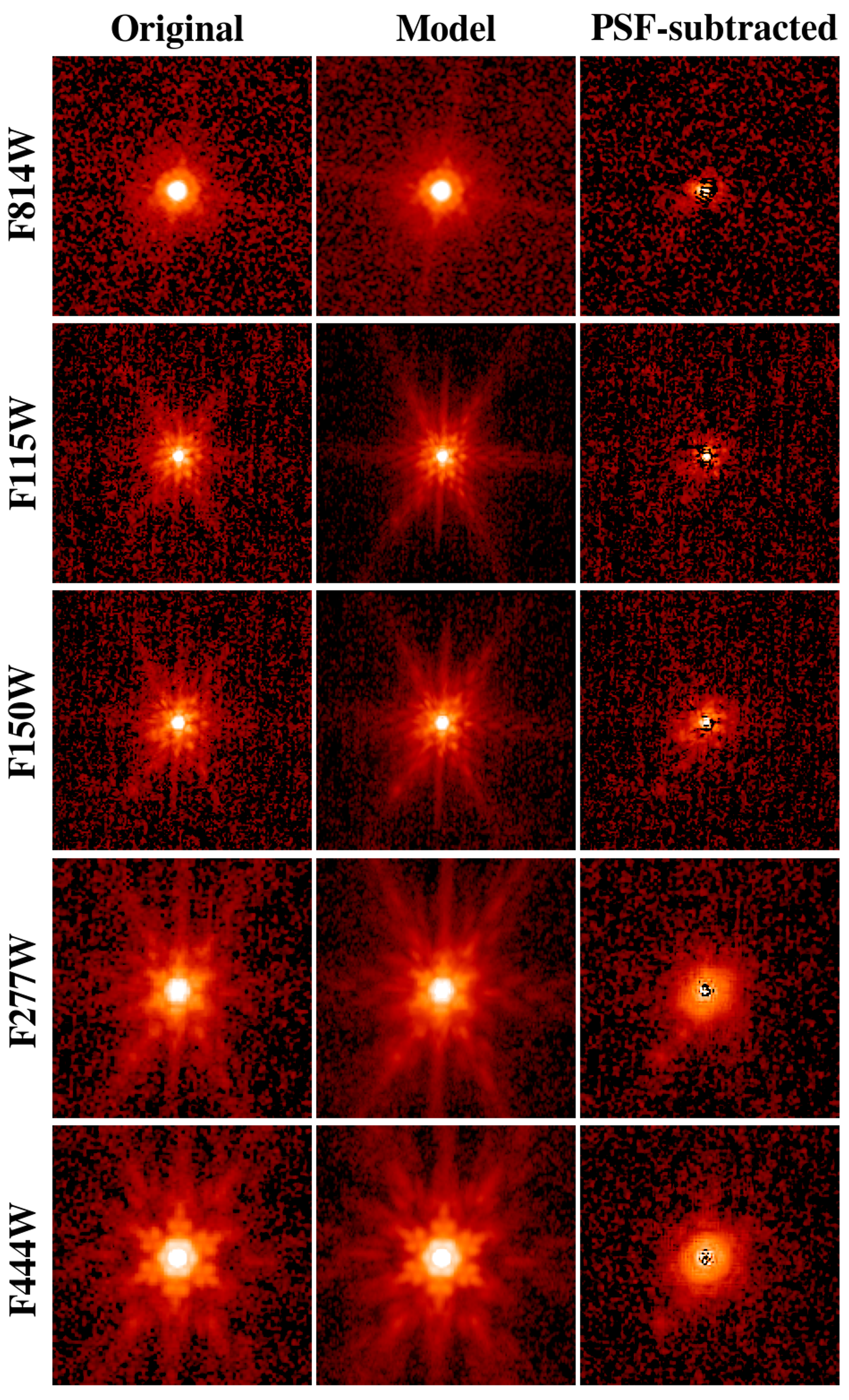}
\caption{This figure shows the NIRCam images from all the observed filters covered in COSMOS-Web campaign. Left panels show the original image, the middle panel shows the image models (see Sect. \ref{sect4.3}) and the right panel shows the residuals after subtracting the models in the middle panel from the original images in the left panel. The presence of two extended companions is clear, especially in the PSF-subtracted images from the filters F150W, F277W and F444W.}
\label{fig:cid346_nircam_allfilters}
\end{figure*}

%%%%%%%%%%%%%%%%%%%%%%%%%%%%%%%%%%%%%%%%%%%%%%%%%%

% Don't change these lines
\bsp	% typesetting comment
\label{lastpage}
\end{document}